\documentclass[12pt]{article}
\usepackage{amsmath,amsthm,amssymb,euscript,epsf,epsfig}
\usepackage{mathtools}
\usepackage{array}
\usepackage{fancybox}
\usepackage{comment} 
\usepackage{wasysym}

\usepackage{hyperref}

\usepackage{rotating}

\usepackage{tikz,pbox}

\usetikzlibrary{shapes,arrows}
\usetikzlibrary{calc,decorations.markings}

\usepackage{circuitikz}
\usetikzlibrary{arrows,decorations.pathreplacing}

\usetikzlibrary{backgrounds,fit,decorations.pathreplacing,calc}

\newcommand{\nnm}{\nonumber}

\makeatletter
\@addtoreset{equation}{section}
\makeatother

\def\one{{\hbox{ 1\kern-.8mm l}}}
\def\zero{{\hbox{ 0\kern-1.5mm 0}}}

\def\mC{ \mathbb{C}}

\def\wchiR{\widehat \chi^R}

\newcommand{\ket}[1]{$|#1\rangle$}

\def\s{ \sigma} 
\newcommand{\Dim}{ {\rm Dim } }

\def\cA{{\cal A}}  \def\cC{{\cal C}}
  \def\cF{{\cal F}}
  
 \def\cK{{\cal K}} 
 \def\cN{{\cal N}} \def\cO{{\cal O}}

 \def\cZ{{\cal Z}}

\def\Rib{ {\rm Rib}}
\newcommand{\id}{\rm id}

\newtheorem{lemma}{Lemma}

\newtheorem{proposition}{Proposition}

\newcommand{\be}{\begin{equation}}
\newcommand{\ee}{\end{equation}}
\newcommand{\beq}{\begin{equation}}
\newcommand{\eeq}{\end{equation}}
\newcommand{\bea}{\begin{eqnarray}\displaystyle}
\newcommand{\eea}{\end{eqnarray}}

\newcommand{\IC}{{\mathbb C}}

\newcommand{\ra}{\rangle}

\newcommand{\cred }{\color{red}}

\newcommand{\bdel}{ {\boldsymbol{\delta}} }

\def\s{ \sigma } 
\def\g { \gamma }

\def\lam { \lambda }
\def\eps { \epsilon }
\def\del { \delta}

\def\reg{ {\rm{reg}} }

\newcommand{\tr}{{\rm tr}}
 
\newcommand{\mR}{\mathbb{R}}

\newcommand{\chim}{ \chi^{\max} }
\newcommand{\chimn}{ \chi^{\max}_{n, k } } 
\newcommand{\chimm}{ \chi^{\max}_{m, k } } 
\newcommand{\chimp}{ \chi^{\max}_{m+n, k} } 

\begin{document}

\begin{flushright}
QMUL-PH-23-04
\end{flushright}

\bigskip

\begin{center}

 
 {\Large \bf   The quantum detection of projectors in finite-dimensional algebras
 and holography
\\
 }
 \medskip

\bigskip

Joseph Ben Geloun$^{a,c,*}$ and Sanjaye Ramgoolam$^{b , d ,\dag}  $

\bigskip

$^a${\em Laboratoire d'Informatique de Paris Nord UMR CNRS 7030} \\
{\em Universit\'e Paris 13, 99, avenue J.-B. Clement,
93430 Villetaneuse, France} \\
\medskip
$^{b}${\em School of Physics and Astronomy} , {\em  Centre for Research in String Theory}\\
{\em Queen Mary University of London, London E1 4NS, United Kingdom }\\
\medskip
$^{c}${\em International Chair in Mathematical Physics
and Applications}\\
{\em ICMPA--UNESCO Chair, 072 B.P. 50  Cotonou, Benin} \\
\medskip
$^{d}${\em  School of Physics and Mandelstam Institute for Theoretical Physics,} \\   
{\em University of Witwatersrand, Wits, 2050, South Africa} \\
\medskip
E-mails:  $^{*}$bengeloun@lipn.univ-paris13.fr,
\quad $^{\dag}$s.ramgoolam@qmul.ac.uk

\begin{abstract} 
We define the computational task of detecting projectors in  finite dimensional associative algebras with a combinatorial basis, labelled by representation theory data, using  combinatorial  central elements in the algebra.  In the first example, the projectors belong to the centre of a symmetric group algebra and are labelled by Young diagrams with a fixed number of boxes $n$.  We describe  a quantum algorithm for the task based on quantum phase estimation (QPE) and obtain estimates of the complexity as a function of $n$.    We  compare to a classical  algorithm related to the projector identification problem by the AdS/CFT correspondence. This gives a concrete proof of concept for classical/quantum comparisons of the complexity of  a detection task,  based in holographic correspondences. A second example involves projectors labelled by  triples of Young diagrams, all having $n$ boxes, with non-vanishing Kronecker coefficient. The task takes as input the projector, and consists of identifying the triple of Young diagrams. In both of the above cases the standard QPE complexities are polynomial in $n$.  A third example of quantum projector detection involves  projectors labelled by a triple of Young diagrams, with $m,n$ and $m+n$ boxes respectively,   such that the associated Littlewood-Richardson coefficient is non-zero. The projector detection task is to identify the triple of Young diagrams associated with the projector which is given as input.  This is motivated by a two-matrix model, related via the AdS/CFT correspondence, to systems of strings attached to giant gravitons. The QPE complexity in this case is polynomial in $m$ and $n$.

\end{abstract}

\end{center}

\noindent  
Key words: quantum information, complexity, ads/cft, tensor models, Kronecker coefficients.

\newpage

\tableofcontents

\section{Introduction and Outlook }

With motivations from the holographic AdS/CFT correspondence \cite{Maldacena:1997re,Gubser:1998bc,Witten:1998qj} and other gauge-string dualities (e.g. \cite{Gross:1993hu,Cordes:1994fc}), we consider the quantum  computational task of detecting projectors in finite dimensional associative algebras. These algebras we consider are semi-simple and have a Wedderburn-Artin (WA) decomposition as a direct sum of matrix algebras (see a general exposition of the WA decomposition in, e.g. \cite{ram}).   They are related to symmetric groups $S_n$ of all permutations of $n$ objects.   The WA decompositions can be constructed using representation theory data from symmetric groups, such as characters of irreducible representations (irreps), matrix elements of permutations in these  these irreps and branching or Clebsch-Gordan coefficients. The projectors of interest act as the identity in a fixed matrix block of the WA decomposition and zero elsewhere. They are thus labelled by the representation theory data specifying the matrix blocks.  Given their relation to symmetric groups, the algebras   also have combinatorial bases related to equivalence classes of permutations. These combinatorial bases are used to construct central elements in the algebras of interest. They are exponentiated to give unitary operators in Hilbert spaces formed by the algebras themselves. Our detection task proceeds with the standard technique of quantum phase estimation (QPE)  \cite{nielsen_chuang_2010}, using these unitary operators as quantum gates which are applied to query the projectors. The set-up is usefully viewed in the context of a quantum communication between Bob, who builds the projector and sends it to Alice, who identifies the projector using QPE. Combining results on query complexities  from quantum information theory and facts from the representation theory of $S_n$, we arrive at complexity estimates which are bounded by polynomials in   $n$.

The AdS/CFT correspondence gives a conjectured equivalence between string theoretic quantum gravity in  
 $AdS_5 \times S^5$ space-time and the non-gravitational quantum theory of $\cN =4$ SYM theory in four dimensions with $U(N)$ gauge group. There are numerous successful tests of the correspondence for interactions of perturbative gravitons (see \cite{Lee:1998bxa} and the review \cite{agmoo}).  Further interesting tests of AdS/CFT  involve the  quantum states associated with half-BPS giant gravitons \cite{McGreevy:2000cw}  and those associated  with half-BPS supergravity geometries (LLM geometries) \cite{Lin:2004nb}. The labelling of half-BPS CFT operators using Young diagrams and the computation of their correlators \cite{Corley:2001zk, Corley:2002mj} plays an important role in the tests involving the interactions of giant graviton branes 
 \cite{Bissi:2011dc,Caputa:2012yj,Lin:2012ey,Kristjansen:2015gpa,Jiang:2019xdz,Yang:2021kot,Chen:2019gsb,Holguin:2022zii} and the dynamics of strings in LLM geometries \cite{deMelloKoch:2020agz}. The construction of the CFT operators labelled by Young diagrams and the computation of their correlators for CFT operators of scaling dimension $n$ uses projection operators in the centre of the group algebra $\mC ( S_n ) $ of the symmetric group $S_n$. The $N$-dependence of the correlators is controlled by the structure constants of the centre 
$\cZ ( \mC ( S_n ) )  $ of the group algebra $\mC ( S_n ) $ \cite{Corley:2001zk}. Recent work 
\cite{Ramgoolam:2023vyq} has shown that by introducing witness fields in matrix models (matrix couplings in the action or classical unintegrated fields in the observables) the structure constants can be exactly reconstructed. 

The correspondence between Young-diagram operators with $ n \sim N^2 $  and half-BPS bulk geometries motivated the formulation of a toy model of information loss \cite{Balasubramanian:2006jt}. It was observed that the multi-pole moments of the space-time fields  correspond to Casimirs of $U(N)$. It was argued that the ten-dimensional Planck scale bounds the observable multipoles. This discussion motivated the consideration in  \cite{Kemp:2019log}  of the problem of detecting general Young diagrams with $n$ boxes using a limited number of Casimirs. The problem was formulated, using 
Schur-Weyl duality, purely in terms of the centre of the symmetric group algebra $\mC ( S_n )$, denoted $ \cZ ( \mC ( S_n ) ) $.  

The simplest and very interesting  Young diagram detection problem in  $\cZ ( \mC ( S_n ))  $, inspired by the holographic physics of half-BPS states in $\cN =4 $ SYM,  is independent of $N$ and concerns the detection of Young diagrams with $n$ boxes. As we explain in section \ref{sec:Background}, $\mC ( S_n) $ and $\cZ ( \mC ( S_n) ) $ are Hilbert spaces.  It is convenient to use language familiar to quantum information theorists and complexity theorists to describe the detection problem. Bob  prepares a quantum state which is a projector $ P_R $ in  $\cZ ( \mC ( S_n ))  $.  He sends the projector $P_R$ to Alice. We will also review the technique of quantum phase estimation for estimating eigenvalues of unitary operators in section \ref{sec:Background}. Alice applies QPE with a set of unitary operators $ \{ U_2 , U_3 , \cdots , U_{ \Lambda (n) } \}$ to determine the label $R$ of the projector. These unitary operators are exponentials of certain Hermitian operators $\{ T_2, T_3 , \cdots , T_{ \Lambda(n) } \}$ which multiplicatively generate $ \cZ ( \mC ( S_n ) )  $.  We will refer to these as cycle central elements.  As we explain in section \ref{sub:centre} choosing the cut-off $\Lambda (n) = n$ suffices to determine $R$, but we expect that a smaller cut-off will suffice. The cut-off is determined by mathematical properties of $\cZ ( \mC ( S_n) ) $ and is equal to   the function $k_* (n)$ identified in \cite{Kemp:2019log}, as we review in section \ref{sec:Background}. 

The symmetric group $S_n$ acts on $n$-fold  tensor product $V_N^{ \otimes n } $ of the fundamental representation of $U(N)$ by permuting the tensor factors. This has been used to construct a basis of half-BPS gauge-invariant operators in $U(N)$ SYM \cite{Corley:2001zk} which is orthogonal in the CFT inner product defined by SYM. These Young diagram operators are related by the  AdS/CFT holographic dictionary to half-BPS supergravity geometries with $AdS_5 \times S^5$ asymptotics \cite{Lin:2004nb}.   The symmetric group algebra operators $T_k$ are related to Casimirs in the $U(N)$ theory, by Schur-Weyl duality, and these have in turn been identified \cite{Balasubramanian:2006jt}  with asymptotic charges of the space-time geometries.  The problem of quantum detection of projectors $P_R$ labelled by one Young diagram of $S_n$ is thus related by holography to a problem of identifying half-BPS geometries with $AdS_5 \times S^5$ asymptotics from their asymptotic multipole moments. We derive  estimates of the complexity of this classical detection in Section  \ref{holog}. It has been recognised that quantum gravity and AdS/CFT present new questions of interest in computational complexity  and quantum gravitational physics which  can have implications for fundamental questions in computational complexity, such as the extended Church-Turing hypothesis \cite{Susskind,Vazirani}. The half-BPS sector of AdS5/CFT4 and its connection to centres of symmetric group
 algebras presents a concrete problem of comparison of complexities for a detection task. The estimates we arrive at in section  \ref{holog} involve $\Lambda (n)$ as well as some intrinsic complexities of gravitational measurements which we leave for future investigation. A more complete treatment of his holographic comparison of complexities  should be useful for other instances of holographic CFT-quantum-state/AdS-gravitational-geometry correspondences (see the recent reference for a detailed discussion involving $AdS_3$ spacetimes e.g. \cite{Bena:2022rna}).  To get some additional perspective on the quantum detection of the projectors $P_R$, we describe a randomized classical algorithm, of the kind used in \cite{TangQC}, for the same task in section \ref{sect:randclass} and give the corresponding complexity estimates.  
 
Section \ref{Kron} considers the problem of detecting projectors in a Kronecker algebra $\cK(n) $ 
\cite{Mattioli:2016eyp,BenGeloun:2017vwn,BenGeloun:2020yau} which is a sub-algebra of $\mC( S_n ) \otimes \mC ( S_n)$. The algebra is isomorphic to a direct sum of matrix algebras labelled by triples of Young diagrams $(R,S,T)$ with $n$ boxes, where these triples have a non-vanishing Kronecker coefficient $C(R,S,T)$.
This is the multiplicity of the trivial representation in the tensor product of $S_n$ irreducible representations $V^{ S_n}_{ R } \otimes V^{ S_n }_{ S } \otimes V^{ S_n}_{ T } $.   The matrix algebras have dimension $C( R, S , T)^2 $. These algebras control the combinatorics and correlators of  invariant observables and correlators of tensor models \cite{BenGeloun:2017vwn}. We imagine a scenario where Bob  constructs the projector and sends it to Alice. Alice can use QPE to identify the triple of Young diagrams in time which is polynomial in $n$. If the triple $(R,S,T) $ has vanishing Kronecker coefficient, there is no projector for Bob to send to Alice. For this reason the projector detection problem falls short of being a quantum algorithm for determining verifying whether or not a triple of Young diagrams has a non-vanishing Kronecker coefficient.  We outline another algorithm, involving communication from Bob to Alice of a simple combinatorial-basis state in $\cK(n)$,  which has an expansion in projectors associated with  all triples $(R,S,T)$ having non-vanishing Kronecker coefficients. Determining whether or not a given triple of Young diagrams appears in this expansion would be a quantum algorithm which determines whether or not the Kronecker coefficient for that triple is zero.  We leave the  problem of estimating the complexity of this algorithm for future research. 


Section \ref{LR} concerns a  third example of projector detection, based on the algebra $ \cA ( m , n ) $ labelled by two integers $m ,n $  and  defined using equivalence classes of permutations in $S_{ m+n}$ modulo conjugation by permutations in $S_m \times S_n$. These algebras arise in the study of 2-matrix invariants relevant to fluctuations of half-BPS giant gravitons in AdS/CFT.   The projectors are labelled by a triple of Young diagrams such that the associated Littlewood-Richardson coefficient is non-zero. The algebra was described in \cite{Mattioli:2016eyp} along with its decomposition into matrix algebras following earlier work on orthogonal bases for CFT4 correlators in the 2-matrix sector \cite{Kimura:2007wy}\cite{Brown:2007xh} \cite{Bhattacharyya:2008rb} \cite{Bhattacharyya:2008xy} \cite{Kimura:2008ac} \cite{Pasukonis:2013ts}  \cite{Kimura:2014mka}.  A formulation of a dual classical computation in terms of strings/branes or gravity configurations is an interesting problem for the future.

Additional future research directions are given at the end of most of the sections. 
Appendix  \ref{app:prooflem}  gives  the proofs of two lemmas of quantum inspired classical algorithms  
used in the main text. Appendix \ref{solvingM} describes a change of basis in the space of Casimirs of $U(N)$ relevant to section \ref{holog}. Appendix \ref{pfpropKron} proves a property of projectors relevant to section \ref{Kron}. 

\vskip.2cm 

\noindent 
{\it Note:} As this paper was approaching completion, we became aware of \cite{BCGH2023} and an associated talk 
 at the Perimeter Institute on the ``Quantum Complexity of Kronecker coefficients". This   has some overlaps with section \ref{Kron} in the use of QPE and projectors to address the computational complexity of Kronecker coefficients.

\section{Background}
\label{sec:Background}

We review some background which will be used in this paper. We start with a review of 
key properties of the group algebra $\mC ( S_n)$ of the symmetric group $S_n$ and its centre $\cZ ( \mC ( S_n ) ) $ which is the subspace that commutes with all elements of $\mC ( S_n)$. This is based on standard textbooks in group theory e.g. \cite{Hamermesh} and the relevant formulae are also collected in Appendices of recent physics papers such as \cite{Pasukonis:2013ts,BenGeloun:2020yau}. This is followed by a review of Quantum Phase Estimation (QPE) \cite{nielsen_chuang_2010} algorithm from quantum computation. 


\subsection{ The group algebra $\mC ( S_n) $  and its centre $\cZ ( \mC ( S_n) )$ }
\label{sub:centre}

Let $\mC ( S_n )$ be the group algebra of the symmetric group $S_n$, i.e. 
the group algebra formed by  linear  combinations $\sum_{\s \in S_n} \lam_\s \s$, $ \lam_\s \in \mC$. 
Let $\cZ ( \mC ( S_n ) ) $ be the center of $\mC ( S_n )$.  There is a basis of $\cZ ( \mC ( S_n ) ) $ 
formed by sums over group elements in conjugacy classes. Conjugacy classes in $S_n$ are labelled by partitions $\mu $ of $n $ (denoted $\mu \vdash n $). For a conjugacy class  $ \cC_{\mu}  \subset  S_n $ 
 we have a central element 
\bea \label{Tmu}
T_{ \mu } = \sum_{ \sigma \in \cC_{ \mu } } \sigma \, . 
\eea
 obeying $\g T_\mu \g^{-1}  = T_\mu$, for any $\g \in S_n$. 
We are interested,  in particular, in conjugacy classes 
 $\mu= [k,1^{n-k}]$ defined by a single cycle of length $k$ 
and all remaining cycles of length 1. These central elements have the form 
\bea 
T_2 &=&   { 1 \over 2 } \sum_{ 1 \le i_1 \ne  i_2 \le n} ( i_1 i_2  ) = (12)+(13)+ \dots    \crcr
T_3 & = &  { 1 \over 3 } \sum_{1\le  i \ne  j \ne  k \le n  } (\, ( i j k ) +(ikj) \,) = (123) + (132) + (124)+ (142) +  \dots  \qquad \cr 
& \vdots &  \cr 
T_k & =&  { 1 \over k  } \sum_{ 1 \ne i_1 \ne i_2 \cdots \ne i_k } ( i_1 i_2 \cdots i_k )  
\eea
where the cycles of length $1$ are not shown. We will refer to the $T_k$ as cycle central elements in 
$ \cZ ( \mC ( S_n )) $. 

The product of two general  central elements for $\mu , \nu \vdash n $  yields 
\bea 
T_{\mu} T_{ \nu } = \sum_{ \lambda } C_{ \mu \nu }^{ \lambda } T_{ \lambda } 
\eea
where the integer structure constants $C_{ \mu \nu}^{ \lambda } $ can be  organized as a matrix $C_{\mu}$ with matrix elements : 
\bea 
( C_{ \mu } )_{ \nu}^{ \lambda } = C_{ \mu \nu}^{ \lambda } \, . 
\eea
It is known that  the eigenvalues of $( C_{ \mu } )_{ \nu}^{ \lambda } $ are the normalized characters $\widehat \chi^R ( T_{ \mu}) :=  { \chi^R ( T_{ \mu} ) \over d_R } $ in the irrep labelled by the Young diagram $R\vdash n$ of dimension  $d_R$.  
 
Another basis for $\cZ ( \mC ( S_n) ) $ is given by the projectors $P_R$ labelled by  irreducible representations $R$ (corresponding to Young diagrams with $n$ boxes, i.e. $R \vdash n $) :  
\bea
P_R = \frac{d_R}{n!}\sum_{\s \in S_n} \chi^{R}(\s) \s
\eea
The  multiplication of the conjugacy class elements with the projectors gives 
\bea
T_{ \mu} P_{ R } 
    = \widehat \chi^R ( T_{ \mu}) P_{ R } 
\eea
where $ \widehat \chi^R ( T_{ \mu}) $ is the normalized character obtained by dividing the trace of the matrix  $D^R ( T_{ \mu } ) $ representing $T_{ \mu } $ in the irrep $R$ by the dimension $d_R$ 
\bea 
\widehat \chi^R ( T_{ \mu}) = { \chi^R ( T_{ \mu } ) \over d_R } ={  \tr ~ (D^R ( T_{ \mu} )) \over d_R } 
\eea
Therefore, the linear operator of multiplication by  $T_\mu$ in $\cZ ( \mC ( S_n) $, with matrix elements $ ( C_{ \mu } )_{ \nu}^{ \lambda } $,  has eigenvectors $P_R$  associated with the eigenvalue
 $\widehat \chi^R ( T_{ \mu})$. 
 We will consider
 \bea\label{maxChi} 
 \chi^{\max}_\mu = \max_{R \vdash n} \widehat \chi^R ( T_{ \mu})
\eea
as a normalization  factor such that
the magnitude of eigenvalue of $T_\mu/  \chi^{\max}_\mu $ associated with $P_R$ 
is bounded by 1, as required by  the QPE procedure.

 The set of central elements  $ \{ T_2 , T_3 , \cdots , T_n \} $ multiplicatively generate the centre, i.e. linear combinations of these and their powers linearly span the centre $\cZ ( \mC ( S_n) ) $, while in fact subsets   $\{ T_2 , \cdots , T_{ k_*(n)} \} $ with $k_*(n) < n $ generically form multiplicative sets of generators \cite{Kemp:2019log}. This is related to the fact that the ordered list of 
of normalized characters $ ( \widehat \chi^{ R } ( T_2  ) , \widehat \chi^{ R }( T_3 ) , \cdots , \widehat \chi^{ R } (T_{k_*(n)}  )  ) $ uniquely identifies the Young diagram $R$. The connection between such ordered lists and sets of multiplicative generators is reviewed and extended to general groups in \cite{Ramgoolam:2022xfk}. The sequence $k_*(n)$   was explicitly computed, with the help of  character formulae in \cite{Lassalle2007AnEF, Corteel2004ContentEA}. Here are some values of $k_* ( n ) $ for lower values of $n$ up to $79$: 
\bea \label{kstar}
k_* ( n )   &  =  & 2 \;  \hbox{ for }   n \in \{ 2,3, 4,5,7 \} \cr 
k_* ( n )   & = &  3  \; \hbox{ for } n \in \{ 6 , 8 , 9 \cdots , 14 \} \cr 
k_* (n) & = &  4\; \hbox{ for } n \in \{ 15 , 16 , \cdots , 23, 25 ,26  \} \cr 
k_{ *} ( n ) &  = &  5 \;\hbox{ for } n \in \{ 24 ,  27 ,   \cdots , 41 \} \cr 
k_* (n)  & = &  6\; \hbox{ for } n \in \{ 42 , \cdots , 78,79  , 81 \} 
\eea

\subsubsection{ A heuristic argument for the asymptotics of $k_* (n)$ at large $n$  } 
\label{heukst} 

A necessary condition for the eigenvalues of   $\{ T_{2} , T_3 , \cdots , T_k \}$ to be able to distinguish all Young diagrams is that the number of distinct lists of  eigenvalues exceeds the number of Young diagrams. The maximum eigenvalue of $T_2$  is $ n (n-1)/2 \sim n^2 $. These eigenvalues are integers. The number of possible eigenvalues is order $n^2$. 
For $T_3$ we have order $n^3$, and $T_k$ order $n^k$ as long as $k << n$. The total number of distinct eigenvalue lists is  approximately 
\bea 
n^2 \times n^3 \cdots  \times n^k = n^{ k  ( k +1 ) /2 }
\eea 
A heuristic estimate for $k_*$ is given by 
\bea 
n^{ k_* (k_* +1 )/2  }   = e^{ \sqrt { n } } 
\eea
Then 
\bea 
k_* ( n ) ( k_* (n )  +1 ) /2 \log n = \sqrt { n } 
\eea
Approximately at large $n $, 
\bea 
k_*(n)  \sim {  n^{ 1/4 } \over  \log n  } \in n^{ 1/4 } 
\eea
Based on this heuristic argument, we conjecture that $k_*(n) \in \cO ( n^{ 1/4} ) $. For our main claim of a polynomial complexity of the QPE-based algorithms we describe subsequently the precise exponent is not important. It suffices that $k_* (n) \in \cO ( n^{ \alpha } )  $ with $ \alpha < 1/2$.

\vskip.2cm

 \noindent{\bf Notation: asymptotic behaviours.}
In the above and henceforth, we use the following notation: for two positive functions
$f,g: \mathbb{N} \to \mR^+$, $f \in \cO(g)$, 
if it exists  a constant $c>0$, and an integer $n_0 \in \mathbb{N} $, 
such that $\forall n\ge n_0$, $f(n) \le  c g(n)$. 
We will equally use, for that relation, the shorthand notation $f \sim g$.
Using a loose language, sometimes, we will say $f$ is $\cO(g)$. 

\vskip.2cm 

\noindent
\subsubsection{ $k_*(n)$, Casimirs of $U(N)$ and multipole moments   } 
\label{kstarLambda} 

This problem of identifying a multiplicatively generating set of cycle central elements of the form $\{ T_2 , T_3 , \cdots , T_{ k_* (n) } \}$ in $ \cZ ( \mC ( S_n) ) $ is related to the problem of identifying Young diagrams with $n $ boxes. There is a transformation giving a sequence of Casimirs of $U(N)$ of the form $\{ C_2 , C_3 , \cdots , C_{k}  \}$ and in terms of a set of cycle central operators $ \{ T_2 , T_3 , \cdots , T_k \}$.  This transformation has been described in the literature on large $N$ 2d Yang Mills theory 
\cite{Cordes:1994fc,Ganor:1994bq}  and has been used to show that elements in the list $\{ C_2 , C_3 , \cdots , C_{k}  \}$ can be expressed in terms of cycle operators with a cut-off at $k$ \cite{Kemp:2019log}.  The work \cite{Balasubramanian:2006jt} relates the Casimirs of $U(N)$ to multi-pole moments of LLM geometries. Therefore the task of distinguishing LLM geometries corresponding to Young diagrams with $n$ boxes can formally be accomplished with a cut-off $\Lambda (n ) $ on the Casimirs with $ \Lambda(n) = k_* (n )$. We 
 are assuming $ N  > n $ where the Young diagram detection problem involves all Young diagrams with $n$ boxes : the $U(N)$ constraint that the first column has length no greater than $N $ is irrelevant. This equality 
 $\Lambda (n ) = k_* (n)$ will be used in section  \ref{holog} where we describe a classical gravity analog of the projector detection task. There is a potential issue in that strictly speaking the validity of the semi-classical super-gravity solution in AdS5/CFT4 requires $n \sim N^2 $. It is possible that this is not an issue for the identification of the highly supersymmetric LLM geometry itself, but only in questions related to fluctuations of the LLM geometry. Another possible scenario for physically  justifying the semi-classical LLM calculation in section \ref{holog} is that the LLM plane may be relevant to a dimensionally reduced model of AdS/CFT, e.g. of the kind considered in the context of spin matrix theory \cite{Harmark:2014mpa}.

\subsubsection{Unitary operators}

We introduce the following operators on $ \mC(S_n)$. 
Consider the linear conjugation operator $S: \mC(S_n) \to \mC(S_n)$ maps a linear combination 
$A = \sum_{i} c_i \s_i \in \mC(S_n)$ to  $S(A) := \sum_{i} c_i \s_i^{-1}$. 
We  easily check that $S^2 = \id$ and call $S$ an involution.  
Let $\bdel: \mC(S_n)^{\times 2} \to \mC$ be the symmetric bilinear pairing 
that acts on  two group algebra elements $A = \sum_{i} c_i \s_i $ and $B  = \sum_{j} c'_j \s_j $, 
as 
\bea
\bdel(A,B) = \sum_{i, j} c_i c'_j    \delta (\s_i  \s_j) 
\eea
with $\delta(s)=1$ if $\s= \id$ and $0$ otherwise. Note that $\delta$ is also extended
by linearity to $\mC(S_n)$ in the following. We then define a sesquilinear form on $\mC(S_n)^{\times 2} $ by 
\bea
g(A,B) = \bdel(\overline{S(A)}, B) ,
\eea
where $\bar x \in \mC$ stands for the complex conjugation. One checks that
$g$ is nondegenerate and therefore induces an inner product  on $\mC(S_n)$. 

We recall a useful proposition and its proof from \cite{BenGeloun:2020yau}. 
\begin{proposition}
For any $\mu \vdash n $, $T_\mu$ is a  Hermitian operator acting on $\mC(S_n)$ endowed with its inner product $g$. 
\end{proposition}
\proof We want to verify the identity  $ g(A , T_\mu B) = g(T_\mu A ,  B)$ for 
$A = \sum_{i} c_i \s_i $ and $B  = \sum_{j} c'_j \s_j $. We have
\bea
&&
 g(A , T_\mu B) = \sum_{i, j} \bar c_i c'_j   \, \del( S(\s_i) T_\mu \s_j)  = 
  \sum_{i, j} \bar c_i c'_j   \, \del( S(T_\mu \s_j) \s_i )  =   \sum_{i, j} \bar c_i c'_j   \, \del( S(\s_j) S(T_\mu) \s_i ) 
  \crcr
 &&
 = \sum_{i, j} \bar c_i c'_j   \, \del( S(\s_j) T_\mu \s_i )  
  =  \sum_{i, j} \bar c_i c'_j   \, \del( S(T_\mu \s_i )\s_j ) =   g(T_\mu A , B) 
\eea
where at an intermediate stage we use $S(AB)= S(B)S(A)$, $S(T_\mu) = T_\mu$, and $\del (S(A))= \del(A)$. 

\qed 

$T_\mu$, as a hermitian operator,  determines a unitary operator $U_\mu = e^{  \frac{2 \pi i}{\chi^{\max}_\mu}   T_\mu } $ with phases  given by the normalized characters $\wchiR(T_\mu)/\chi^{\max}_\mu$: $U_\mu = e^{  \frac{2 \pi i}{\chi^{\max}_\mu}    T_\mu } P_R 
= e^{  \frac{2 \pi i}{\chi^{\max}_\mu}    \wchiR(T_\mu) } P_R $. 
It is therefore natural to use QPE to estimate the phases of such unitaries. 
The next task uses QPE with  C-$U_\mu$-gates  
to identify the corresponding normalized characters $\wchiR ( T_\mu ) $ and from that addresses new questions.

\subsection{ Quantum Phase Estimation} 
\label{sec:QPEreview}

We review in this section the QPE algorithm and its rum time complexity. 
We use   Nielsen and Chuang's notation, see section 5.2 in \cite{nielsen_chuang_2010}. 
Readers with a background in quantum information theory may wish to skip this section and simply glance at the last paragraph of this section. Those with an AdS/CFT and mathematical physics background but little quantum information, like ourselves at the start of this project, will hopefully find it useful.

Consider  a unitary operator $U$ with an eigenstate $|\psi \ra$ of eigenvalue $e^{2\pi i \lambda} $, 
for an unknown $\lambda$.  QPE is a quantum algorithm that aims at
approximating the value of $\lambda$. It uses  
 oracles (black boxes) giving access to controlled-$U^{2^j}$ ($C\!-\!U^{2^j}$ or shortly CU) operations, for some $j$,
  and capable of preparing an initial state $|\psi \ra$. 
As in many quantum algorithms, the fast Fourier transform defines
a key subroutine of QPE. Finally the last important step in the procedure
is a measurement that will deliver an approximation of the phase with a 
defined error rate (tolerance).  Figure \ref{fig1} gives the standard quantum  
circuit that provides the different phases of the QPE. 
 In the following we use a register space of $t$ qbits, i.e. the dimension of that space is 
 $2^{t}$.  Without loss of generality, the eigenvalue $\lambda$ is chosen such that $\lambda\in [0,1]$. 
For our applications of QPE in the upcoming sections, it will be the case that $\lambda $ has an exact $t$-bit expansion, in which case, as we see shortly,  the exact eigenvalue can be determined by QPE  with probability $1$. 

%

\begin{figure}
\begin{center}
\begin{tikzpicture}[thick]
\tikzset{meter/.append style={draw, inner sep=5, rectangle, font=\vphantom{A}, minimum width=18, line width=.8,
 path picture={\draw[black] ([shift={(.1,.3)}]path picture bounding box.south west) to[bend left=50] ([shift={(-.1,.3)}]path picture bounding box.south east);\draw[black,-latex] ([shift={(0,.1)}]path picture bounding box.south) -- ([shift={(.3,-.1)}]path picture bounding box.north);}}}
    \tikzstyle{operator} = [draw, fill=white, minimum size=1.5em] 
      \tikzstyle{qft} = [draw, fill=white, minimum height = 9em, minimum width = 3em] 
    \tikzstyle{phase} = [draw, fill, shape=circle, minimum size=5pt, inner sep=0pt]
    %
    \matrix[row sep=0.2cm, column sep=0.5cm] (circuit) {
      \node (q5) {\ket{0}}; &    \node[operator] (H51) {H}; &    &  &   \node[] (P54) {\dots};  &  \node[phase] (P55) {};   & 
        \coordinate (end5); & &   \coordinate(E56) ;  & \coordinate(E57){}; & \node[meter](meter) at (0,0) {} ;  & \coordinate(end5f)  ; \\
    \node (q3) {\vdots}; &    \node[] (H31) {}; &   & &    \node[] (P34) {};  &     &   \coordinate (end3);  
   & &   \coordinate(E36) ;  &  &  & \\
 \node (q2) {\ket{0}}; &    \node[operator] (H21) {H}; &  &    \node[phase] (P23) {};   &    \node[] (P24) {\dots};  
 &   &     \coordinate (end2); & &   \coordinate(E26) ;  & \coordinate(E27){}; & \node[meter](meter2) at (0,0) {} ;  & \coordinate(end2f)  ;  \\
    \node (q1) {\ket{0}}; & [-0.5cm]  \node[operator] (H11) {H}; &   \node[phase] (P12) {}; &   &    \node[] (P14) {\dots}; & &[-0.3cm]     \coordinate (end1);  & &   \coordinate(E16) ;  & \coordinate(E17){}; & \node[meter](meter1) at (0,0) {} ;  & \coordinate(end1f)  ;  \\
     \node (q6) { \ket{\psi} }; &    &   \node[operator] (U62) {$U^{2^0}$}; 
     &   \node[operator] (U63) {$U^{2^1}$};  &   \node[] (U64) {\dots};  &  \node[operator] (U65) {$U^{2^{t-1}}$}; &   
     \coordinate (end6); & & \coordinate(E66); &   \coordinate(E67); &    \coordinate (end6f); \\
    };

   %
        \draw[decorate,decoration={brace, mirror},thick]
        ($(circuit.north west)-(0.3cm,0.3cm)$)
       to node[midway,right] (bracket) 
       {}  ($(circuit.south west)+(-0.3cm,1cm)$)  ;
%
       
    \begin{pgfonlayer}{background}
            
        \draw[thick]   ($(end5) +(0cm,0.3cm)$) -- ($(end1)+ (0cm, -0.3cm)$) -- ($(end1) +(0.98cm,-0.3cm)$) --  ($(end5) +(0.98cm,0.3cm)$) -- ($(end5) +(0cm,0.3cm)$) ;
        \draw[thick] (q1) -- (P14) -- (end1)  (q2) -- (P24) --(end2)  (q5) -- (P54) -- (P55) -- (end5)
                               (U62) -- (P12)  (U63) -- (P23) (U65)--(P55) (q6) -- (U64) -- (end6) 
                                (E56)--(E57)--(meter)--(end5f)
                                (E26)--(E27)--(meter2)--(end2f) 
                                (E16)--(E17)--(meter1)--(end1f) 
                                 (U65) -- (end6f) ;
    \end{pgfonlayer}
    %
    \end{tikzpicture}
    \put(-410,76){\scriptsize{1st register}}
        \put(-410,63){\scriptsize{$t$-qbits}}
         \put(-410,10){\scriptsize{2nd register $\;\;\;\,\Big\{$}}
         \put(-93,76){\scriptsize{QFT$^{-1}$}}
\end{center}
\caption{Quantum phase estimation by a quantum circuit acting on the initial state $|0 \ra^{\otimes t} \otimes | \psi \ra $: 
 H-boxes are Hadamard gates, $U^{2^i}$-boxes stand for CU-operators, $i=0,\dots, t-1$, $QFT^{-1}$ for the inverse quantum Fourier transform, 
and the last stage involves a measurement on the first register. } 
\label{fig1}
\end{figure}
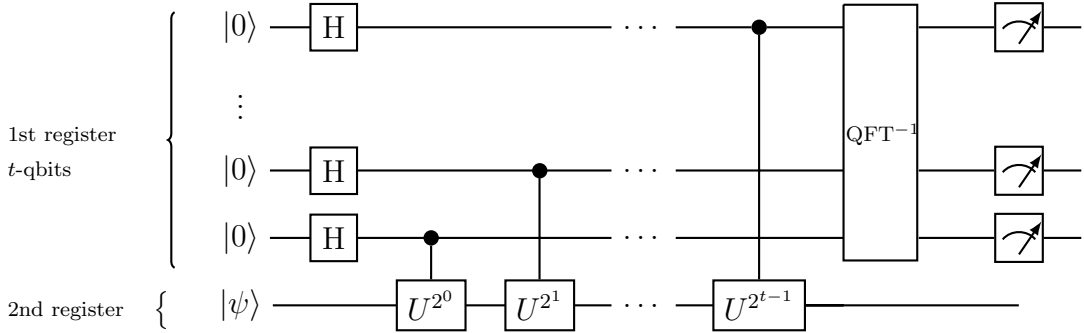

\begin{figure}
\begin{center}
\begin{tikzpicture}[thick]
    \tikzstyle{operator} = [draw, fill=white, minimum size=1em] 
    \tikzstyle{phase} = [draw, fill, shape=circle, minimum size=3pt, inner sep=0pt]
    \matrix[row sep=0.2cm, column sep=0.2cm] (circuit) {
      \node (q1) {\ket{j_1}}; &    \node[operator] (H11) {H}; &  \node[operator] (R12) {$R_2$};  
       &    \node[] (P13) {\dots}; &   \node[operator] (R14) {$R_{t-1}$};  & \node[operator] (R15) {$R_{t}$};   
        &    &  & &   &     & & & & \coordinate(end1f)  ; \\
    \node (q2) {\ket{j_2}}; &                                                     &   \node[phase] (P22) {};   
      &    \node[] (P23) {\dots}; &    &       &    \node[operator] (H26) {H};     &   \node[operator] (R27) {$R_2$};  &   \node[] (P28) {\dots};    &  \node[operator] (R29) {$R_{t-2}$};   &\node[operator] (R210) {$R_{t-1}$} ;  & && &\coordinate(end2f)  ;  \\
    \node (q3) {\vdots}; &    \node[] (H31) {\vdots}; &   &   \node[] (P33) {\dots}; &    &     &      & &   &  &  & && &\\
      \node (q4) {\ket{j_{t-1}}}; & &   &    \node[] (P43) {\dots}; &  \node[phase] (P44) {}; & &   & &   & \node[phase] (P49) {};  &    &  \node[operator] (H410) {H}; &  \node[operator] (R411) {$R_2$};  & &\coordinate(end4f)  ;  \\
      
     \node (q5) { \ket{j_{t}} }; &    &      &    \node[] (P53) {\dots};  &   &  \node[phase] (P55) {}; &   & &&  & \node[phase] (P510) {};  & & \node[phase] (P511) {}; &\node[operator] (H512) {H};  &\coordinate (end5f); \\
    };
     
    \begin{pgfonlayer}{background}
    
        \draw[thick]  (R12) -- (P22)    (q1) -- (P13)  (P13)-- (end1f) (q2) -- (P28)  (P28)-- (end2f)    
                              (q4) -- (P43) (P43) --  (end4f) 
                              (q5) -- (P53) (P53) -- (end5f) 
                              (R29) -- (P49)
                              (R14) -- (P44) 
                              (R15) -- (P55)
                              (R210) -- (P510)
                              (R411) -- (P511); 
    \end{pgfonlayer}
    \end{tikzpicture}
\end{center}
\caption{A circuit for quantum Fourier transform $|j_1j_2 \dots j_{t} \ra $: 
 H-boxes are Hadamard gates, $R_k$-boxes stand for C-$R_k$-operators, $k=2,\dots, t$.} 
\label{fig2}
\end{figure}
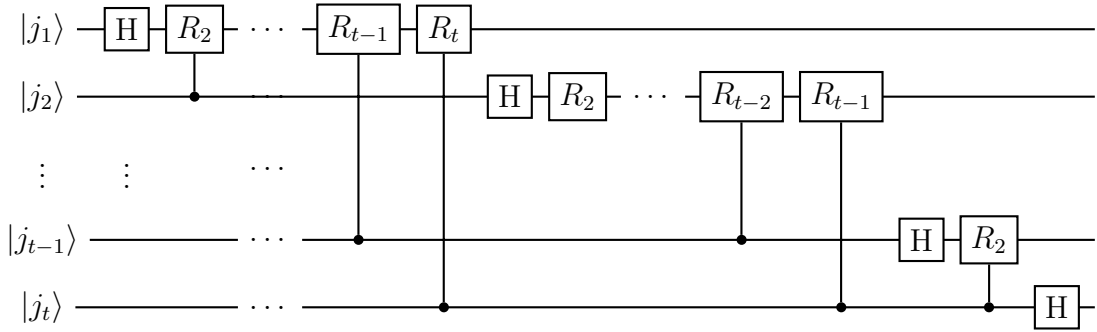

 
QPE algorithm goes through the following steps: 

(1) Initialize  a state $|0\ra^{\otimes t} \otimes |\psi \ra  $, apply the Hadamard-gates (H-gates) on the first register
to get  
\bea
| \psi \ra_ 0  = 
\frac{1}{ 2^{t/2}} ( | 0  \ra +  |1 \ra ) ^{ \otimes t} | \psi \ra  
 = \frac{1}{ 2^{t/2}} \sum_{j=0}^{2^t - 1} | j \ra  |  \psi \ra  
\eea
where  $| j \rangle$ is (the computational basis) written in binary representation as $ | j_1j_2 \dots j_{t} \rangle$ 
and $j = \sum_{l=1}^{t} j_l 2^{t-l}$. 
Recall that the Hadamard operator is expressed by 
$H = |+\rangle \langle 0| + |-\rangle \langle 1|$ where $|\pm \rangle = 
(1/\sqrt{2}) (\, |0\rangle \pm   |1\rangle \,)$. 

(2) Apply the CU-gates (evolution operator): 
\bea
|\psi_1 \rangle  =
  \Big[\bigotimes_{i=1}^t  C\!-\!U^{2 ^i } \Big]| \psi \ra_ 0  =
\frac{1}{ 2^{t/2}}   \sum_{j=0}^{2^t-1}  
e^{2 \pi i j   \lambda } | j \rangle | \psi \rangle 
\eea
where each CU-gate is defined by $C\!-\!U^{2 ^i }   ( | 0  \ra +  |1 \ra )| \psi \ra  
=   | 0  \ra | \psi \ra  +   |1 \ra U^{2 ^i }  | \psi \ra $.

(3) Apply the inverse quantum Fourier transform (QFT$^{-1}$) to the first register of $|\psi_1 \rangle$, with C-$R_k$-operators defined by 
\bea
R_k = \left(\begin{array}{cc}
1 & 0\\
0 & e^{2\pi i /2^{k} }
\end{array}\right)
\eea
and obtain 
\bea
|\psi_2 \rangle  &=& \frac{1}{ 2^{t}}  \sum_{l=0}^{2^t-1}
 \sum_{j=0}^{2^t -1}  e^{  2 \pi i j ( \lambda - \frac{l}{2^t} )} | l \rangle |  \psi \rangle 
 \eea
 Two cases may occur: 
 
 (a) suppose that $\lambda$ has  an exact $t$-bit expansion,  
there exists an $l^*$ such that $l^*/2^t$ coincides
with $ \lambda$. The above expression takes the form
 \bea
 |\psi_2 \rangle 
 &  = &  | l^* \rangle |  \psi \rangle 
 + 
  \frac{1}{ 2^{t}}  \sum_{l=0|  l \ne l^* }^{2^t-1} 
  \sum_{j=0}^{2^t -1}  e^{  2 \pi i j ( \lambda - \frac{l}{2^t} )} | l \rangle |  \psi \rangle 
   \crcr
 &  = &
  | l^* \rangle |  \psi \rangle 
 +  \frac{1}{ 2^{t}}  \sum_{l=0 |  l \ne l^* }^{2^t-1} 
  \Big[\frac{1 - e^{ 2 \pi i ( 2^t  \lambda -  l  ) } }{ 1 -e^{ 2 \pi i ( \lambda - \frac{l}{2^t} )} } \Big] | l \rangle |  \psi \rangle \crcr
 &   =& 
   | l^* \rangle |  \psi \rangle 
 +  \frac{1}{ 2^{t}}  \sum_{l=0 |  l \ne l^* }^{2^t-1} 
  \Big[\frac{1 - e^{ 2 \pi i (l^* -  l  ) } }{ 1 -e^{ 2 \pi i ( \lambda - \frac{l}{2^t} )} } \Big] | l \rangle |  \psi \rangle
  \label{finalqpe}
\eea
 the second term identically vanishes and we obtain
\bea
|\psi_2 \rangle  =   | l^* \rangle |  \psi \rangle
\eea
\label{prob1} 

(4) Perform measurements of the $t$-bits in the first register.  
This yields  $\lam = l^*$  in binary form with probability 1. 
  
  \

(b)  If $\lam$ does not have an exact $t$-bit expansion, then 
consider the best $t$-bit estimate $b$ of $\lambda$: 
$b \in [\![0, 2^t -1]\!]$  is the largest integer such that 
$b/2^t = b_1 2^{-1}+ b_2 2^{-2} + \dots + b_{t}2^{-t} < \lambda$.  
We define $\zeta =  \lambda - b/2^t $, $0<\zeta< 2^{-t}$, and write 
\bea
|\psi_2 \rangle & =& \frac{1}{ 2^{t}}  \sum_{l=0}^{2^t-1}
 \sum_{j=0}^{2^t -1}  e^{  2 \pi i j ( \lambda - \frac{l}{2^t} )} | l \rangle |  \psi \rangle
 = \frac{1}{ 2^{t}}  \sum_{l=0}^{2^t-1}
 \sum_{j=0}^{2^t -1}  e^{  2 \pi i j ( \zeta - \frac{(l-b)}{2^t} )} | l \rangle |  \psi \rangle  
\crcr
&=&
 \frac{1}{ 2^{t}}  \sum_{l=0 }^{2^t-1} 
  \Big[\frac{1 - e^{ 2 \pi i ( 2^t \zeta  - ( l-b)  ) } }{ 1 -e^{ 2 \pi i (  \zeta - \frac{l-b}{2^t} )} } \Big] | l \rangle |  \psi \rangle \crcr
  &   = &
 \frac{1}{ 2^{t}}
 \sum_{l=-b }^{2^t-1-b} 
     \Bigg[\frac{1 - e^{ 2 \pi i ( 2^t \zeta -  l  ) } }{ 1 -e^{ 2 \pi i (  \zeta - \frac{l}{2^t} )} } \Bigg] | l +b\rangle |  \psi \rangle
 \eea
 Nielsen and Chuang mapped the amplitude of $| l +b\rangle |  \psi \rangle$
 to $\alpha_l \in \IC$ defined in the following way: 
 \bea
 &&
 \alpha_l \; \text{is the amplitude of the state} \; | l +b\rangle \;  \text{ labelled by}\;  l= 0,1,\dots, 2^t -1- b \\
 &&
 \alpha_{2^t +l} \;\text{is the amplitude of the state} \; | l +b\rangle \;  \text{ labelled by}\;\;  l= -b,-b+1,\dots,-1  \nonumber
 \eea
It is shown that $| \alpha_l| \le 1/[2^{t+1} (\zeta  -\frac{l}{2^t} )]$
that will be useful to achieve the next probability bound. 

($4^{\prime } $) Perform measurements of the $t$-bits in the first register.  
It is  proved (see section 5.2 in \cite{nielsen_chuang_2010}) that the probability of getting $b$ after a measurement 
$m$ obeys the probability bound $P(|m-b|> e) < \eps = 1/(2(e-1))$, for a parameter 
$e = 2^{t-p}-1= 2^t \eta - 1$. Thus, if $p$ is much smaller than $t$, $e$ becomes
large and induces a  probability $P(|m-b| \le e) \ge  1- \eps$ close to $1$. 

\

\noindent{\bf Remarks.} 
Although our present work solves a task 
that takes as input the eigenvectors  of an evolution 
operator (which are projectors in an algebra), similar quantum algorithms to what we will present hereafter 
address other important questions and take superpositions of states as input (see discussion of the first future direction in section \ref{Kron}). Then, QPE can be implemented with any linear combination of eigenstates
such as $| \Psi \ra = \sum_{k=1}^n \beta_k | \psi_k \ra $, for the eigenbasis  $\{\psi_k\}$
 of a Hermitian operator, and for coefficients $\beta_k \in \IC$. 
Following step by step the previous algorithm, the end result takes the form
 \bea
|\psi_2 \rangle  = \frac{1}{ 2^{t/2}}    \sum_{l=0}^{2^t-1} \sum_{k=1}^n \beta_k 
 \sum_{j=0}^{2^t-1}  e^{  2 \pi i j  (\lambda_k  - \frac{l}{2^t})} | l \rangle | u_k \rangle 
 \eea
Consider a given  eigenvalue $\lam_k$ approximated by the fraction $b_k/2^t$. 
  After measurement,   we obtain a probability $P(|m -b_k|> e) < |\beta_k|^2 \eps $
 that  remains small, for constant $\beta_k$. 
  Thus, the  procedures  which follow here  will adapt  in this case and will 
 remain accurate.

\



\

\noindent{\bf Complexity.}
The runtime complexities of QPE has been also discussed in \cite{nielsen_chuang_2010}. 
The query complexity is $\cO(t)$, as one requires $t$ queries of the CU-operators. 
On the other hand, the gate complexity, that enumerates the number of 
gates used to deliver an output, is $\cO(t^2)$ ($t$ Hadamard gates, $t$ CU-operators, 
the inverse QFT  requires $t + (t-1) + \dots +2+1$ $H$ and C-$R_k$ gates, 
and therefore asymptotically leads to $\cO(t^2)$ gates). 
Here and after,  we summarize these statements as: 

{\bf Query complexity}$\,\in \cO( t)$; 
 
{\bf Gate complexity}$\,\in \cO( t^2)$. 

In the projector detector tasks for algebras related to $ S_n$ which we consider in the following sections, where we will use unitary operators  for QPE, which are exponentials of the cycle central operators discussed in section \ref{sub:centre},  we will be choosing the number of bits $t$ to allow exact determination of the relevant eigenvalues which have known integrality properties. As a result this will translate into $n$-dependent complexities which are bounded by polynomials in $n$, 
, subject to the assumption that $k_*(n) \in \cO ( n^{ \alpha } ) $ with $ \alpha < 1/2$ which we described in  section \ref{heukst}.

\ 

\section{Quantum Phase Estimation for projectors in the centre of $\mC ( S_n) $}
\label{sect:qpesn}

In this section, we focus on a quantum algorithm for detecting projectors
in the centre  $\cZ ( \mC ( S_n)) $  of the group algebra $\mC ( S_n)$  of the symmetric group $S_n$.
Bob prepares a projector $P_R$ as a quantum state in the Hilbert space $\cZ ( \mC ( S_n)) $. 
Alice is tasked with identifying the projector, i.e identifying the Young diagram $R$. 
Her procedure uses QPE to obtain the eigenvalues of appropriate Hermitian operators (the cycle central elements 
described in section  \ref{sub:centre}) acting on the  Hilbert space.  These eigenvalues are expressible in terms of normalized characters of irreducible representations of $S_n$.   We estimate the query and gate complexities of the algorithm using their standard definitions.


We are interested in particular  in the cycle central elements $T_k$,  $k\in [\![1,n ]\!]$, corresponding to partitions $\mu = [k, 1^{n-k}]$ with one cycle of length $k$ and remaining cycles of length $1$, described in more detail in section \ref{sub:centre}. 
Consider $U_k = e^{  \frac{2 \pi i}{\chi^{\max}_k }   T_k } $ and 
$ U_k P_R = 
e^{  \frac{2 \pi i}{\chi^{\max}_k }   \widehat {\chi}^R ( T_k ) }  P_R$, using the definition \eqref{maxChi} for $\chi^{\max}_k $.  Alice's undertaking is the following : 

\begin{center}
{\bf TASK: } Given the set of unitaries $U_2, U_3, \dots, U_k$, $k\in [ \! [2, n  ]\! ]$, and state $P_R$ with unknown $R\vdash n$, determine which $P_R$ we have.
\end{center}
We may imagine that Bob prepares the state $P_R $ in $ \cZ ( \mC ( S_n ) ) $ and sends it to Alice who is tasked with the identification. 

To answer this question, we will use QPE with $C-U_k$-gates. We will subsequently compare with 
standard complexities of  classical algorithms. 

\subsection{QPE for  $T_2$}

We restrict to this simplest (and non-trivial) case for a warm-up. 
We will use the known fact that \cite{lasalle}
\bea 
\wchiR(T_2) =
{ \chi^R ( T_2 ) \over d_R } = \sum_{ \Box \in R  } c ( \Box ) \, . 
\eea
The sum is over boxes in the Young diagram $R$, weighted by the content of the box. The content is the column number minus the row number.
The maximum value of
the normalised character occurs for $ R = [ n ]$, the Young diagram with a single row of $n$ boxes. 
Then we have 
\bea 
\sum_{ i =1}^n ( i-1) = { n ( n -1 ) \over 2 } 
\eea
The lowest value is $ - n ( n -1) /2$ and corresponds to $R=[1^n]$. Take for example $ n =6$, then $R= [6]$ gives $\wchiR(T_2) = 6*5/2 = 15$;  for $ R = [ 3,2,1]$, one has $ { \chi^R ( T_2 ) \over d_R } = 0$. For $ R = [ 2,2,2] $ it is  $-3$, for $ R = [ 3,3]$ it is $5$ ; for $ R = [4,2] $ it is $5$. For simplicity, considering the value of $\wchiR(T_2)$, let us allow for all the integers between the bounds 
$ - n ( n -1) /2$ and $ n ( n -1) /2$.

The register must have enough bits to be able to store in binary all the numbers from 
$ -  n ( n -1) /2$ to $ n ( n -1)/2$.  This means the register has to be able to store $ n ( n -1) +1 $ possible integers. The number of bits needed
is given by 
\bea 
t = \lceil \log_2 ( n ( n-1) +2 ) \rceil \in \cO(\log n)
\label{tT2}
\eea

Thus, according to \cite{nielsen_chuang_2010} (see Chapter 5.2), the QPE has complexities: 

{\bf Query complexity}$\,\in \cO( t) = \cO(\log (n)) $; 
 
{\bf Gate complexity}$\,\in \cO( t^2) = \cO (( \log (n) )^2)$. 

\ 

We now give an algorithm that will detect with probability 1, the projector
$P_R$ with a register of $t = \log n$ bits.  $QPE(U_2)$ stands for 
the QPE with $C-U_{2}$-operators. 
The query and gate complexities are the same as above.

\subsection{General procedure for arbitrary $n$}


\vskip.2cm 

From our previous discussion (particularly equation \eqref{prob1} and the associated remark (4), 
we  know that QPE can be done with 
certainty provided the eigenvalue can be exactly expressed with a $t$-bits. 
For instance, if $ n \in \{ 2,3,4, 5,  7 \}$, $T_2$ determines the projector $P_R$. 
 However, for higher $n$, characterizing $P_R$ will require access to more
operators $T_k$'s as generating the centre $\cZ ( \mC ( S_n ) )$ requires more of these operators,
see the discussion in subsection \ref{sub:centre}. 

For $n \in \{ 6 , 8, \dots, 14 \}$, we need $T_2 , T_3$. Given a single copy of $P_R$,  we can 
separately estimate the eigenvalues of $T_2$ and $T_3$. To measure the eigenvalue of $T_2$, in the exact way 
as above, Alice applies  QPE for $U_2 = e^{  \frac{2 \pi i}{\chi^{\max}_2 }   T_2 } $ using conditional unitaries
 C-$U_2$.   Using \eqref{prob1} the  output of the first stage is $P_R$, tensored with some ancillary bits which are measured 
to give the eigenvalue $\wchiR(T_2)$ of $T_2$. 
Then we apply QPE, using  $U_3 = e^{  \frac{2 \pi i}{\chi^{\max}_3 }    T_3} $  and corresponding 
conditional unitaries  C-$U_3$-gates. The measure will give with certainty $\wchiR(T_3)$ of $T_3$.  The  data pair $(\wchiR(T_2),\wchiR(T_3))$ determines in a unique way $P_R$ by searching $R$ in a table of size $2*|{\rm partitions}(n)|$. 
 
For larger $n$, we simply iterate the procedure, enlarging the dimension of the register space 
and introducing a larger number of $T_k$, $k \in [\![2,n]\!]$, that surely determine $P_R$. 
We need to perform the analysis at generic $T_k$, for $k \in [\![2,n]\!]$. 

\subsubsection{Dimension of the register space for eigenvalues of general $T_k$ }

A first task is to collect the dimension of the register space
associated with the calculation of the projector eigenvalues. 
We use a  result on character bounds by Feray and Sniady \cite{F_ray_2011}. 
They tell us that 
\bea 
| { \chi_{ R} ( T_{ \mu } ) \over d_R } | \le a  |T_{\mu} | [  {\rm max } ( { r( R ) \over n } , { c ( R ) \over n } , { |\mu| \over n }      ) ]^{ |\mu | } 
\eea
where $ a$ is a constant $a > 0 $; $ |\mu| $ is the minimum number of factors needed to write a permutation in $T_{ \mu} $ as a product of transpositions; $ r ( R ) $ is the number of rows; $c(R)$ is the number of columns in $R$.
We observe that 
\bea 
{\rm max } ( { r( R ) \over n } , { c ( R ) \over n } , { |\mu| \over n }      ) 
\le {\rm max } ( { n  \over n } , { n  \over n } , { |\mu| \over n }      ) \le 1 
\eea
using the fact that the number of rows and number of columns cannot exceed $n$. 
Thus, we have a bound 
\bea
| { \chi_{ R} ( T_{ \mu } ) \over d_R } | \le a  |T_{\mu} | 
\eea
Dealing with $\mu = [ k , 1^{ n -k} ] $, i.e. the case of interest
associated with the cycle central elements $T_k$, we obtain the asymptotic behavior, for fixed $k $, 
\bea
| { \chi_{ R} ( T_{ \mu } ) \over d_R } | &\le& a  {  n ( n-1) \cdots ( n - k+1) \over k } 
 = \frac {a n^k}{k}  (1- \frac1n)(1 -\frac2n ) \cdots (1 - \frac{k-1}{n}) \crcr
&\le&  \frac {a n^k}{k}  \Big(1 -  \sum_{l=1}^{k-1} \frac{l}{n} + \hbox{ terms vanishing  at large $n$}  \Big) 
\crcr
&\le& \frac {a n^k}{k} \Big( 1 - \frac{k(k-1)}{2n} + \hbox{ terms vanishing at large $n$ }   \Big)
\eea
We are assuming that the cut-off $k_*(n) $ on the range of $k$ needed obeys $k_*(n) <<  n$, which is certainly compatible with the conjecture $k_*(n) \in \cO ( n^{ 1/4}  ) $ explained in section \ref{heukst} but $k_*(n) \in \cO ( n^{ \alpha } ) $ for $\alpha < 1 /2 $ suffices to ensure that the second and higher terms in the expansion above are sub-leading at large $n$. 

We also know \cite{Simon1995RepresentationsOF}  that $ { \chi_{ R} ( T_{ \mu } ) \over d_R } $  (for any partition $\mu$) is an integer.  So this means that the range of possible values of these eigenvalues is $ 2a n^k$ and the smallest difference between eigenvalues is $1$.
Thus, on very general grounds, we can conclude that  an upper bound on the number of bits requires to encode all eigenvalues of  the C-$U_k$ operators  is given by 
\bea
\hbox{ Bits required to store eigenvalues of $T_k$  } &\sim& \log_2 ( 2  \frac {a n^{k}}{k}   ) 
\\
&\sim& \log_2 ( n^k ) - \log_2 k  \sim  k \log n - \log k 
\nonumber
\eea
%
%


\subsubsection{Query and gate complexities at general $n$}

We will need to add  the complexities for the measurements of $T_2, T_3 , \cdots , T_{ k_*(n) } $. 
The total query complexity is estimated as 
\bea
&&
\sum_{ k =2}^{ k_* (n) }   k \log (n ) = ( {  k_* (n) ( k_*(n) +1 )\over 2 } - 1 )  \log (n )    \crcr
&&
\sum_{ k =2}^{ k_* (n) }  ( k \log (n )  - \log (k) ) =  {  k_* (n) ( k_*(n) +1 )\over 2 } \log (n )   - \log( k_*(n)!)  \crcr
&&
\sim  {  k_* (n) ( k_*(n) +1 )\over 2 } \log (n )   - k_*(n) \log( k_*(n))
\eea
If   $ k_* (n) < n^{ \alpha } $ holds at large $n$, we have 
\be \label{querysum}
t \sim  {  k_* (n) ( k_*(n) +1 )\over 2 } \log (n )   - k_*(n) \log( k_*(n))
\sim  n^{\alpha } (n^{\alpha } -  \alpha) \log n
\sim  n^{2\alpha}  \log n
\ee
Thus, we obtain the 

\noindent{\bf Query complexity} $ \in \cO(t) \sim \cO(n^{2\alpha} \log n)$. 

For the gate complexity, we sum the number of gates required for each step and estimate
\bea 
\sum_{ k =2}^{ k_* (n) }   k^2 ( \log (n ) )^2  \sim {  k_* (n) ( k_*(n) +1 ) ( 2 k_* (n) +1 ) \over 6 } ( \log (n ) )^2    
\eea
If we use $k_*(n) = n^{ \alpha } $ at large $n$, we obtain 

\noindent{\bf Gate complexity} $ \in  \cO(n^{3\alpha} (\log n)^2 )$. 
According to the heuristic conjecture in section \ref{heukst}, $\alpha = 1/4$, but for any $\alpha < 1/2$, the above gate and query complexities hold. 

%
%
%
%
%

\section{Classical algorithm related to quantum projector identification by holographic duality} 
\label{holog} 

Projectors in symmetric group algebras  $\mC ( S_n) $ are used to construct half-BPS operators in $ \cN =4$ super-Yang Mills theory with $U(N)$ gauge group, which is dual to string theory on $AdS_5 \times S^5$.   These are composite operators built from $n$ copies of a  complex matrix $Z$. First we define 
\bea 
\cO_{ \sigma } ( Z ) = \sum_{ i_1 , \cdots , i_n }  Z^{ i_1 }_{ i_{ \sigma (1)} } \cdots Z^{ i_n }_{ i_{ \sigma (n)} }
\eea
This definition can be extended, by linearity, to include operators labelled by group algebra elements : 
\bea 
\cO_{ \sum_{ \sigma } c_{ \sigma } \sigma } = 
\sum_{ \sigma } c_{ \sigma } \cO_{ \sigma} ( Z ) 
\eea
It then follows that for projectors 
\bea 
P_{ R } = { d_{ R } \over n! } \sum_{ \sigma } \chi^R ( \sigma ) \sigma 
\eea
we have operators 
\bea 
\cO_{R } = { d_R \over n! }  \sum_{ \sigma } \chi^R ( \sigma ) \cO_{ \sigma  } ( Z ) 
\eea
Up to an unimportant normalization these are the standard Schur polynomial operators of the half-BPS Sector, which form an orthogonal basis for the free field inner product \cite{Corley:2001zk}.  These states can be realized as holomorphic operators created by complex matrix oscillators arising from the $U(N)$  $\cN=4$ SYM theory reduced on $S^3 \times \mR$.   The dynamics of the reduced matrix model is isomorphic to that of a many-body system of $N$ fermions in a one-dimensional  harmonic oscillator potential \cite{Corley:2001zk, Berenstein:2004kk, Takayama:2005yq}.  Lin, Lunin and Maldacena \cite{Lin:2004nb}  classified the  half-BPS super-gravity solutions with $AdS_5 \times S^5$ asymptotics, which take the form:
 \bea 
 ds^2 = - h^{-2} ( dt + \sum_{ i =1}^2 V_i dx_i )^2 + h^2 ( dy^2 + \sum_{ i =1}^2 dx_i dx_i ) + R^2 d \Omega^2 + R^2 d \tilde \Omega^2  
 \eea
The functions $V_1 , V_2 , h , R $ appearing above are all functions of $ (x_1 , x_2 , y )$, and are all determined by one function $u ( x_1 , x_2 , y )$. The function obeys a harmonic equation in $y$ and is determined by its value on the $y=0$ plane.

In  the papers   \cite{Balasubramanian:2005mg, Balasubramanian:2006jt}, the function $u ( x_1 , x_2 )$ on the LLM plane is determined by using a Wigner phase space distribution associated to the quantum many-body fermion state. This Wigner distribution arises from a reduced one-particle density matrix obtained from the many-body state by tracing out the states of $N-1$ particles.  The upshot of this discussion is  eq.\,(25) from  \cite{Balasubramanian:2006jt} which gives  an expression 
for $u$ that allows the determination of the conserved charges from the semi-classical geometry : 
\bea \label{urhotet}
u ( \rho , \theta ) 
=  2 \cos^2 \theta \sum_{ l =0}^{ \infty } { \sum_{ f\in \cF } A^{ l } ( f ) \over \rho^{ 2l +2 } } 
 (-1)^l ( l +1 )\;  {}_2F_{1}  ( -l , l +1 ; 1 ; \sin^2 \theta ) 
\eea
Here  $\rho\in [0,\infty]$, $\theta\in [0, \frac{\pi}{2}]$, 
$\cF = \{f_1, f_2, \dots, f_N\}$ a set of increasing integers related
to the eigenvalues of individual Fermion $E_i = \hbar (f_i + \frac12)$, 
$i=1,2,\dots, N$. In such an expansion, 
$A^{ l } ( f ) $ is a polynomial of order $l$ in $f$ 
(its explicit form can be found in \cite{Balasubramanian:2006jt}).

The relation between Jacobi polynomials and hypergeometrics will be useful 
(see \cite{CHEN1995153}) 
\bea 
P_n^{\alpha , \beta } ( x )  =
\begin{pmatrix}  \alpha + n \cr n  \end{pmatrix}  
{}_2 F_{ 1} (  -n , \alpha + \beta + n +1 ; \alpha +1 ; { 1 - x \over 2  } ) 
\label{jacobipoly}
\eea
Observe that we can match the hyper-geometric to the Jacobi polynomials by using 
\bea 
&& \alpha = 0 \,, \qquad   \beta = 0 \,, \qquad   n = l  \crcr
&& { 1 - x \over 2 }= \sin^2 \theta   \implies x = \cos ( 2 \theta  ) 
\eea
Then, under such a restriction, $u ( \rho , \theta )$ \eqref{urhotet} becomes 
\bea 
u ( \rho , \theta )  = 2 \cos^2 \theta \sum_{ l =0}^{ \infty } {\sum_{ f\in \cF }  A^{ l } ( f ) \over \rho^{ 2l +2 } }  (-1)^l ( l +1 )  
P_{ l }^{ 0 , 0 } ( \cos 2 \theta ) 
\eea
Further useful information on Jacobi polynomials, including orthogonality relations, are given in  \cite{CHEN1995153}.

Introducing, $X = \cos (2\theta)$, and a cut off $\Lambda$ in $l$, we  write an approximation of $u(\rho,\theta)$ at large enough  $\Lambda$ as 
\bea\label{urhoX}
u(\rho,X, \Lambda) = (1+X) \sum_{ l =0}^{ \Lambda }
U(l,\rho) P_{ l }^{ 0 , 0 } ( X ) 
\eea
where $U(l,\rho)$ given by 
\bea 
 U(l,\rho)  =  {\sum_{ f\in \cF } A^{ l } ( f ) \over \rho^{ 2l +2 } }  (-1)^l ( l +1 ) 
\eea  
The cut-off $\Lambda (n)$ is the cut-off on $U(N)$  Casimirs (for $N > n $ ) needed to ensure that the Young diagrams can be distinguished. As explained in section \ref{kstarLambda} it can be identified with 
   $k_* (n)$ which is defined intrinsically in terms of the symmetric group algebras   
 $\mC ( S_n) $ and their centres $\cZ ( \mC ( S_n ) ) $.  
The Casimirs of interest are given  by 
\bea
\sum_{ f\in \cF } A^{ l } ( f ) 
 = (-1)^l  \rho^{2l +2}\,  \frac{ U(l,\rho) }{l+1}
\eea
 and therefore can be extracted once we know 
 $ U(l,\rho)$. 
We  define a  reduced form of the function $u$ by the equation 
\bea
\widetilde u  (\rho,X, \Lambda)
= \frac{u(\rho,X, \Lambda)}{ (1+X) } = 
\sum_{ l =0}^{ \Lambda } U(l,\rho) P_{ l }^{ 0 , 0 } ( X )
\eea
and this will be used in place of $u$. 

\

\noindent
{\bf  Fast Fourier Transform (FFT) for $\widetilde u(\rho,\theta, \Lambda)$.} 
We use a well-established method,  the FFT, to extract
the coefficients $U(l,\rho) $ of $\widetilde u(\rho,\theta, \Lambda)$. Our first task is to put $\widetilde u(\rho,\theta, \Lambda)$ in a 
convenient form to apply the FFT algorithm.

 We use the  expansion of the Jacobi polynomial
where $ p_{ l, k }^{ 0 , 0 }$ is the coefficient
in the expansion defined by the series \eqref{jacobipoly}, see \cite{CHEN1995153}: 
\bea
&&
P_{ l }^{ 0 , 0 } ( \cos 2\theta ) =
 \sum_{k=0}^l  p_{ l, k }^{ 0 , 0 }\;  ( \cos 2\theta ) ^k 
 \crcr
 &&
 = \sum_{k=0}^l  p_{ l, k }^{ 0 , 0 } \; 
  \frac{1}{2^k}\sum_{j=0}^k 
  \binom{k}{j} e^{2i\theta (2j - k)} 
   = \sum_{k=0}^l  p_{ l, k }^{ 0 , 0 } \; 
  \frac{1}{2^k}\sum_{m=-k |\;  m+k \text{ even }}^{ k}
  \binom{k}{\frac{k+m}{2}} e^{2i\theta m } 
    \crcr
  &&
   = \sum_{m=- l }^{l}
   \Big[  \sum_{k=|m| |\;  m+k \text{ even }}^l    \frac{ p_{ l, k }^{ 0 , 0 } }{2^k} 
   \binom{k}{\frac{k+m}{2}}  \Big] \; e^{2i\theta m }
    =  \sum_{m=- l }^{l}    \widetilde p_{l,m} 
     \; e^{2i\theta m }
    \crcr
    && 
   \widetilde p_{l,m} := 
\sum_{k=|m| |\;  m+k \text{ even }}^l    \frac{ p_{ l, k }^{ 0 , 0 } }{2^k} 
   \binom{k}{\frac{k+m}{2}} 
   \label{jacteta}
 \eea 
 In the following, we will require access to the coefficients   $\widetilde p_{l,m} $ of the Jacobi polynomial 
 $P^{0,0}_l$. Thus we use as input  data table 
 $\{ \widetilde p_{l,m},\;  l=0,1,\dots, \Lambda, \; 
 m=-l, -l+1,\dots, l \}$.

Using \eqref{jacteta}, we reorganise the sum 
 $\widetilde u(\rho,\theta, \Lambda)$  in the following fashion
 \bea\label{utilde}
 &&
\widetilde u(\rho,\theta, \Lambda)
=  
\sum_{l=0}^{ \Lambda } U(l,\rho)
\sum_{m=- l }^{l}
     \widetilde p_{l,m}   \; e^{2i\theta m } 
\crcr
&&
= \sum_{m=- \Lambda }^{\Lambda }
\Big[
\sum_{l=|m|}^{ \Lambda }U(l,\rho)\; 
   \widetilde p_{l,m}     \Big] e^{2i\theta m } 
= 
 \sum_{m=- \Lambda }^{\Lambda }
 \widetilde C_{m}  (\rho,  \Lambda)\;  e^{2i\theta m }  
\crcr
&&
\widetilde C_{m} (\rho,  \Lambda) := 
\sum_{l=|m|}^{ \Lambda }U(l,\rho) \; 
   \widetilde p_{l,m}  
\eea
One can  easily show that $\widetilde C_{m}
 (\rho, \Lambda) = \widetilde C_{-m} (\rho, \Lambda)$ which allows to return to a real expansion in $\cos(2 m \theta)$. 
 We focus  on the values of 
  $\widetilde C_{m} (\rho,  \Lambda)$,   for $m=0,\dots, \Lambda$. 
Suppose  that we only have access to 
a finite number $\Lambda+1$ of values of $\widetilde u(\rho, \theta_l = \frac{\pi l}{\Lambda+1} ,  \Lambda )$, $l=0,\dots, \Lambda$.

Then, the discrete Fourier transform (DFT) of $ \widetilde u(\rho, \theta ,  \Lambda )$ 
 delivers the coefficients $\widetilde C_{m}(\rho,  \Lambda) $: 
\bea
&&
\frac{1}{\Lambda+1} \sum_{l=0}^{\Lambda} \widetilde u( \rho,  \frac{\pi l}{\Lambda+1},  \Lambda) 
(e^{- \frac{2i\pi l }{\Lambda+1}  } )^m  \crcr
&& = \frac{1}{\Lambda+1}
\sum_{l=0}^{\Lambda} 
 \sum_{m'=- \Lambda }^{\Lambda }
 \widetilde C_{m'}  (\rho,  \Lambda) \;  e^{2i \frac{\pi l}{\Lambda+1} (m' -m)} 
 \crcr
 &&
 = \frac{1}{\Lambda+1}
\sum_{l=0}^{\Lambda} 
\Big[ \widetilde C_{m}  (\rho,  \Lambda) 
+ 
 \sum_{m'=- \Lambda | m'\neq m }^{\Lambda }
 \widetilde C_{m'}  (\rho,  \Lambda) \;  e^{2i \frac{\pi l}{\Lambda+1} (m' -m)} \Big]\crcr
&&
= 
 \widetilde C_{m}  (\rho,  \Lambda)  + 
 \frac{1}{\Lambda+1}
  \sum_{m'=- \Lambda | m'\neq m}^{\Lambda }
   \widetilde C_{m'}  (\rho,  \Lambda)
  \frac{1- e^{2i \frac{\pi  (\Lambda+1)}{\Lambda+1} (m' -m)} }{1 - e^{2i \frac{\pi }{\Lambda+1} (m' -m)} }
  \crcr
  &&
 =   \widetilde C_{m}  (\rho,  \Lambda) 
\eea
The DFT has computational complexity  $\cO(\Lambda^2)$: we have order $\Lambda$ coefficients to extract and
order $\Lambda$ multiplications (and additions of 
smaller complexity) to perform. 
Rather than doing a DFT, 
we perform a FFT to extract the coefficients $\widetilde C_{m}(\rho,  \Lambda)  $ of 
 $\widetilde u  (\rho, \theta, \Lambda)$.  This is a classical algorithm, 
based on  decimation-in-time, 
with $\Lambda = 2^r$ points, see e.g. 
\cite{Heckbert1998FourierTA}, 
FFT decomposes the DFT in $r= \log_2 \Lambda$ 
steps, each of which generating $\Lambda/2$ basic computations
called butterflies. 
In the end,   FFT  costs $\cO(\Lambda \log 
\Lambda)$ basic operations.


The FFT delivers a table of values of $ \widetilde C_{m}  (\rho,  \Lambda) $, $m=0,\dots , \Lambda$. 
We can then invert a linear system to obtain 
the $U(l,\rho)$: 
\bea
&&
 \widetilde C_{\Lambda}  (\rho,  \Lambda) 
 =U(\Lambda,\rho)    \widetilde p_{ \Lambda,  \Lambda}
    \qquad 
    \qquad 
    U(\Lambda,\rho)  = \frac{1}{   \widetilde p_{ \Lambda,  \Lambda}} \widetilde C_{\Lambda}  (\rho,  \Lambda) \crcr
 &&
 \widetilde  C_{ \Lambda-1}  (\rho,  \Lambda) =
 \sum_{l=\Lambda-1}^{ \Lambda }U(l,\rho)
  \widetilde p_{l ,  \Lambda-1}
  = 
   U(\Lambda-1,\rho)
     \widetilde p_{\Lambda-1 ,  \Lambda-1}
    + U(\Lambda,\rho) \widetilde p_{\Lambda ,  \Lambda-1}
\cr\cr
&& 
    U(\Lambda-1,\rho)  = 
    \frac{1}{ \widetilde  p_{ \Lambda-1, \Lambda-1 } }
    \Big[   \widetilde  C_{ \Lambda-1}  (\rho,  \Lambda)  
    -  U(\Lambda,\rho)  
     \widetilde  p_{ \Lambda, \Lambda-1 } \Big]
   \cr\cr
   &&
     U(j ,\rho) 
     = 
    \frac{1}{ \widetilde  p_{ j, j } }   \Big[  
       \widetilde  C_{j}  (\rho,  \Lambda)  
        - 
     \sum_{l=j+1 }^{ \Lambda }U(l,\rho)   \widetilde  p_{ l, j } \Big] \crcr
   &&
   j = \Lambda-1,  \Lambda- 2, \dots, 0
   \label{Usolv}
\eea
We use the fact that the 
$k$th coefficient $p_{l, k }^{ 0 , 0 } $ of the Jacobi polynomial $P^{0,0}_l$
expresses as (expand equation (1.1) of \cite{CHEN1995153}): 
\bea
p_{l,k}^{0,0} =   \frac{(-1)^{k}}{l!}   \sum_{m=k}^l  \, \binom{l}{m} \, 
\binom{m}{k}  
\frac{ (l+m)! }{2^m m! }  (-1)^{m}  
\eea
and thus  the diagonal coefficient  $\widetilde p_{l,l}  = \frac{(2l)!}{2^l ( l !)^2}$ which is inverted in \eqref{Usolv} does not vanish.

The above procedure accesses  the  table of values  $\{\widetilde p_{l,m}\}$, and depends on a parameter $ \Lambda (n)$ which obeys $\Lambda (n) \in \cO (  n^{ \alpha } )$. 
We assume that the computational complexity of 
measuring the $\widetilde u(\rho, \theta_l = \frac{\pi l}{\Lambda+1} ,  \Lambda )$, $l=0,\dots, \Lambda,$
is  bounded from above by a certain function 
$c_{\widetilde u}(\Lambda)$, the complexity of measuring $\widetilde  u$ at separations of $2\pi/(\Lambda+1)$. The estimation of $c_{\widetilde u}(\Lambda)$ will require a complexity analysis of measurements in classical gravity, which we leave for future discussion and calculation. 

\ 

Two  procedures are involved above in  the determination
of the coefficients $U(l,\rho)$:  a FFT and the resolution of the 
triangular linear system. 
The FFT  has  computational 
cost  $\cO(\Lambda \log \Lambda)$ 
and inverting this  system 
costs $\cO(\Lambda^2)$. Combining this
yields a complexity $f(\Lambda)$ bounded 
from above by 
\bea
f(\Lambda)  \le  
c_0 \; \Lambda\;  c_{\widetilde u}(\Lambda) + 
c_1 \Lambda \log \Lambda +
c_2 \Lambda^2   
\eea 
where $c_0, c_1 $ and $c_2$ are constants. 
Hence, two sub cases could occur: 

\quad  {\bf Case 1: }  if 
$c_{\widetilde u}(\Lambda)\in \cO ( \Lambda^{ \beta } )  $ with $\beta \le 1$, 
for large enough $ \Lambda$, then the overall cost is $f(\Lambda) \in  \cO( \Lambda^2 )$; 

\quad 
{\bf Case 2: } if $c_{\widetilde u}(\Lambda) \in \cO ( \Lambda^{ \beta } ) $ with $\beta >1 $ 
for large enough $ \Lambda$ and for some constant $c_0>0$, then $f(\Lambda)$ will be dominated by the cost of the measurement procedure $ \cO(\Lambda^{ 1 + \beta } )$.

In  first case 1.1, we have the FFT and linear system inversion that are combined to give a complexity $f ( \Lambda ) \in  \cO ( \Lambda^2 ) \sim \cO ( n^{ 2 \alpha })  $. Comparing to the QPE  query and gate complexities  given by  $ \cO(n^{2\alpha}\log n)$,
and $ \cO(n^{3\alpha} (\log n)^2)$, respectively, the classical holographic procedure has a better  efficiency
at large $n$. 
In case 1.2, if $1<\beta \le 2$, 
QPE query complexity is smaller  than the 
holographic complexity, 
but the QPE gate complexity is larger. 
If $\beta >2$, QPE has both 
query and gate complexities smaller 
than the holographic complexity.

It would be interesting to see if  
 a more refined analysis would distinguish concepts of query and gate complexities during the determination of $U$ in the holographic setting.  There is another subtlety which is worth mentioning.  It concerns the inversion of the linear system which may bring a limitation  
of the above conclusion.  The coefficients $\widetilde p_{l,m}$,
for some large  $l$ and $m$, 
may have a size comparable to $\Lambda^\kappa$, for
some positive integer $\kappa$. 
Therefore, we can use the fast Schonhage-Strassen multiplication or division by  $\widetilde p_{l,m}$ 
that costs 
$\cO(\Lambda^\kappa \log \Lambda)$  \cite{Harvey2021} whenever we perform  $U(l,\rho)   \widetilde  p_{ l, j }$ 
or divide by $ \widetilde  p_{ j , j }$. 
A more detailed analysis of $\kappa$ would be needed to 
improve the comparison  between the complexity of detection of projector  quantum states  and the complexity of detection of geometries in the holographic dual. We leave these refinements for the future. The calculations in the present paper should be viewed as a proof of concept for the existence of meaningful comparisons of complexity  between classical gravitational algorithms based on  AdS and quantum algorithms based on CFT which are made possible by known holographic dualities. This is broadly  aligned with the perspectives in \cite{Susskind,Vazirani}. 
 
 
 
 
%
%

Note finally that one could be interested
 in solving the power sum moments $M_k= \sum_{i=1}^N f_i^k$. This can be also performed because
of the relation
$ \sum_{ f\in \cF } A^{ l } ( f ) =  \sum_{k=0}^l  c^l_k M_k $,
for some coefficients $c^l_k$, see Appendix \ref{solvingM}. 
These relations can be recast as a triangular linear system
that can be solved. For this reason, we take the view that 
 knowing the $ \sum_{ f\in \cF } A^{ l } ( f ) $ is equivalent to knowing the $M_k$. 

 
 \vskip.2cm

 \noindent 
{\bf  Future direction: Finite $N$  effects.  } 

\vskip.1cm

In the above discussion, we have considered the complexity of the problem of detecting a
Young diagram with $n$ boxes from the set of all Young diagrams with $n$ boxes. In $\cN =4$ SYM CFT this corresponds to considering the Young diagram basis for operators of dimension $n$. We have been able to formulate well-defined complexity estimation problems using the form of the function $ u ( \rho , \theta )$ determining the LLM solutions. 
An interesting extension of the complexity estimates given here is to consider the space of Young diagrams with $N^2$ boxes, for a large integer $N$, restricted to have the first column of length no greater than $N$. This is motivated by the condition of validity of the super-gravity approximation in describing the physics of high dimension CFT operators.

\section{Classical matrix algorithms for projector detection: randomised classical algorithm } 
\label{sect:randclass}

In section \ref{holog} we described a classical algorithm which related by  holography to the 
 quantum detection of projectors in section  \ref{sect:qpesn}. A more conventional approach to classical/quantum comparison of algorithms is to used quantum-inspired randomised classical algorithms \cite{TangQC}.  
In this section we will develop such an approach for the projector detection task.

We can compare classical and quantum algorithms for this problem of detecting projector states. In the classical problem, we are given a set of matrices of size $D$, one matrix corresponding to every projector $P_R$. We can construct such matrices by taking the group algebra elements 
\bea 
P_R = { d_R \over n! } \sum_{ \sigma } \chi^R ( \sigma ) \sigma
\eea
in a reducible representation of dimension $D$. We need a reducible representation which, in its decomposition into irreducibles, contain every irrep $R$ at least once. 

We will work with the regular representation.

\

\noindent 
{\bf  {\bf TASK 1 }  : The task of finding the $T_2$ eigenvalue  }

 \noindent 
 The input 
\begin{itemize} 

\item We are given, as a black-box which we can query, 
a  projector $P_R$ as a matrix $D^{ V } ( P_R ) $ of size $\Dim V \times \Dim V$. 

\item We are also given, as a black box, a matrix $ D^V ( T_2 )$.

\end{itemize}

\ 

\noindent 
{ \bf Algorithm} 

\begin{itemize} 

\item We query the $D^{ V } ( P_R ) $ black box and ask for a pair of numbers  $(i,j)$ 
$ 1 \le  i , j \le \Dim V $ with a non-zero entry $ (D^V ( P_R) )_{ij} $. 

\item We consider the matrix equation 
\bea\label{fixedmatTP}  
( D^V ( T_2)  D^V ( P_R) )_{ij} = \sum_{ k } (  D^V ( T_2 ) )_{ik}  (D^V ( P_R))_{ kj }   =
{ \chi^R (T_2) \over d_R   } ( D^V ( P_R) )_{ ij} 
\eea 
We query the $i$'th row $\{ (D^V ( T_2))_{ik} : k \in \{ 1, \cdots , \Dim V \} \}   $ of the matrix $D^V ( T_2) $ and the $j$'th column $\{ (D^V ( P_R))_{ kj} : k \in \{ 1, \cdots , \Dim V \}  \}  $ of $D^V ( P_R)$. 

\item The sum over $k$ in the equation \eqref{fixedmatTP} is  computing  a scalar product of two vectors. The two vectors are real. We will do the calculation of this inner product using a randomized classical algorithm ($l_2$-norm sampling as in \cite{TangQC}). This will lead to a complexity which is independent of $\Dim V$ and instead depends on the admissible error. This step is detailed below. 

\item By taking the ratio of $  ( D^V ( T_2)  D^V ( P_R) )_{ij}          $ over $(D^{ V } ( P_R ))_{ij}$ we can get the eigenvalue.

\end{itemize} 

We now turn to the key computational step above is the multiplication of the $i$'th row of $D^V ( T_2) $ with the $j$'th column $(D^V ( P_R)) $. This is the computation of a dot-product of two vectors in dimension $\dim V $. We can immediately apply known results in quantum-inspired classical algorithms . 
We use the result in Proposition 4.2 of \cite{TangQC},  which gives the number of queries required to calculate the inner product of two vectors $< x,y> $ with error $\epsilon || x || || y|| $, and with success probability $ ( 1 - \delta))$. 
The number of queries needed is $O ( {1 \over \epsilon^2} \log ( {1 \over \delta} ) )$. Here we are calculating $ (T_2 P_R)_{ ij}$ which is 
$ {\chi^R ( T_2 )/d_R } $ times $(D^V ( P_R))_{ ij}$. We know the range of the  normalized characters extends within  the interval $ [ - n^2 , n^2 ] $ 
  and we know they are integers.  From this we extract an estimate for the biggest $\epsilon$ we can afford. This will lead to query complexity which is polynomial in $n$, as we now show.

\

Let us consider the application of Proposition 4.2 of \cite{TangQC} to projectors in the regular representation. 

Some important lemmas (to be proved in Appendix \ref{app:prooflem}) : 
\begin{lemma} \label{lem1}
\bea 
\sum_{ \gamma } ( D^{ \reg}_{ \gamma \mu }  ( P_R ) )^2 = { d_{R}^2 \over n! } 
\eea
\end{lemma}
\begin{lemma}  \label{lem2}
\bea 
\sum_{ \tau } ( D^{ \reg}_{ \sigma \tau  } ( T_2 ) )^2  = { n ( n-1) \over 2} 
\eea
\end{lemma}
Useful to choose fixed $ \sigma , \mu $ with the property that $D^{ \reg}_{ \sigma \mu }  ( P_R ) \ne 0 $. Then define vectors of dimension $n!$ 
\bea 
&&  X_{ \gamma }  =  D^{ \reg}_{ \gamma \mu }  ( P_R ) \cr 
 && Y_{ \tau} =   D^{ \reg}_{ \sigma \tau  } ( T_2 )
 \eea
 The above lemmas are telling us 
 \bea 
&&  ||X||^2 = { d_R^2 \over n! } \cr 
 && ||Y||^2 = { n ( n -1) \over 2 } 
 \eea
 Also we have 
 \bea 
 \langle X , Y \rangle  && = \sum_{ \gamma } (  D^{ \reg}_{ \sigma \gamma   } ( T_2 )( D^{ \reg}_{ \gamma \mu }  ( P_R )\cr 
 && = D^{ \reg}_{ \sigma \mu }  ( T_2 P_R ) \cr 
 && = { \chi^R ( T_2 ) \over d_R }  D^{ \reg}_{ \sigma \mu }  (  P_R )
 \eea
Lemma 4.2 of \cite{TangQC} tells us that error 
\bea 
\Delta ( \langle  X , Y \rangle ) = \epsilon ||X|| ||Y|| 
\eea
can  be achieved with query complexity $ O ( { 1 \over \epsilon^2} \log \frac{1}{\delta} ) $, where $\delta$ is  the failure probability to get within
$ \epsilon ||X|| ||Y|| $ of $ \langle X , Y \rangle$. 
 In the following, we neglect $\log \frac{1}{\delta}$ as it is 
 independent of $n$.  
In the present application, we are taking $ D^{ \reg}_{ \sigma \mu }  (  P_R )$ to be given to us by an oracle, so it is an exactly  known quantity. So we have  
\bea 
\Delta (  \langle X , Y \rangle )  = \Delta (     { \chi^R ( T_2 ) \over d_R }    )
 D^{ \reg}_{ \sigma \mu }  (  P_R )
\eea
Equivalently, dividing by the known quantity $ D^{ \reg}_{ \sigma \mu }  (  P_R )$ we have 
\bea
\Delta (  {  \langle X , Y \rangle  \over   D^{ \reg}_{ \sigma \mu }  (  P_R ) }   ) 
= \Delta (     { \chi^R ( T_2 ) \over d_R }    ) 
\eea
We want this uncertainty to be less than $1$ : then we can use the integrality property of the normalised characters to ensure that we can use the classical algorithm to detect the normalized character. If we set 
\bea 
\Delta (     { \chi^R ( T_2 ) \over d_R }    )  = \epsilon ||X|| ||Y|| \le  1 
\eea
we get 
\bea 
\epsilon \le { 1 \over ||X|| ||Y|| }  =   \sqrt { n! \over d_R^2 } .  \sqrt{ 2 \over n (  n -1) }  \sim 
\sqrt {n! \over d_R^2 } { 1 \over n } 
\eea
Let us call the upper bound on $ \epsilon$ as $\epsilon^* $. So we have 
\bea 
\epsilon^* = \sqrt {n! \over d_R^2 } { 1 \over n} 
\eea
This leads to a lower bound on the query complexity which we call $Q^*$
\bea 
Q^* ( R )  = { 1 \over ( \epsilon^* )^2} = n^2 . { d_R^2 \over (n!) } 
\eea
This complexity is maximised for the largest $d_R$. 
There are easily derived bounds on the $d_R^{ max} $
\bea 
\label{dRmax}
\sqrt{ \frac{|G| - a(G)}{c(G) - a(G)} } \le d_R^{\max} \le \sqrt{|G| - a(G)}.
\eea 
as discussed in math-overflow \cite{373786}. Here $ |G| = n! $ , $ c(G)$ is the number of conjugacy classes $p(n)$. $a(G)$ is the number of irreps of dimension $1$.  The only irreps having dimension $1$ are the trivial and the anti-symmetric, so $a(G) = 2$. So we get 
\bea\label{bds}  
 { n! -2     \over   (p(n) - 2 )  } \le (d_R^{\max})^2  \le  ( n! -2 )  
\eea
We conclude 
\bea\label{ClassCompT2}  
Q^* ( R_{ max} ) \le n^2 . { ( n! - 2) \over (n!) } \sim n^2 
\eea

\ 
%
%
%
 
 \noindent 
{\bf {\bf TASK }  2 : The task of finding the $T_k$ eigenvalue  for general $ k \in \{ 2 ,  3, \cdots , k_* ( n ) \} $   }

 The input 
\begin{itemize} 

\item We are given, as black-box which we can query, 
a  projector $P_R$ as a matrix $D^{ V } ( P_R ) $ of size $\Dim V \times \Dim V$. 

\item We are also given, as  black boxes, the matrices $ D^V ( T_2 )$, 
$ D^V ( T_3 )$, \dots, $ D^V ( T_{k_*(n)} )$

\end{itemize} 

We apply the same previous algorithm successively for  $ T_2, T_3 , \cdots ,$ and $T_{ k_*(n) } $. 

We use the matrix of $T_k$ in the regular representation, i.e. $D^{ \reg}(T_k)$
that should fulfil the following relation. 

\begin{lemma}  \label{lemk}
\bea 
\sum_{ \tau } ( D^{ \reg}_{ \sigma \tau  } ( T_k ) )^2  = { n ( n-1)\cdots (n-k+1) \over k} 
\eea
\end{lemma}
\proof 
Use the same steps as in proof of Lemma \ref{lem2}  in 
Appendix \ref{app:prooflem}, the above sum 
boils down to  $\delta(T_k^2) = \sum_{c, c' \in \cC_k} \delta(\s_c \s_{c'})
= |T_k| = n!/(k(n-k)!)$, where $\cC_k = \{\sigma \in  [k, 1^{n-k}]\}$
contains all permutation with cycle structure given by 
$ [k, 1^{n-k}] $. 

\qed 

Using the regular representation, with $D^{ \reg}_{ \s \mu }  ( P_R ) \ne 0$, 
at step $k$, we define vectors of dimension  $n!$ 
\bea 
  X_{ \gamma }  =  D^{ \reg}_{ \gamma \mu }  ( P_R ) \,, \qquad 
 Y_{ \tau} =   D^{ \reg}_{ \sigma \tau  } ( T_k)
 \eea
 Lemmas \ref{lem1} and \ref{lemk} shows us that
 \bea 
  ||X||^2 = { d_R^2 \over n! }  \,, \qquad    ||Y||^2 = { n ( n-1)\cdots (n-k+1) \over k} 
 \eea
Computing the inner product, we obtain 
 \bea 
 \langle X , Y \rangle  = \sum_{ \gamma }   D^{ \reg}_{ \sigma \gamma   } ( T_k ) D^{ \reg}_{ \gamma \mu }  ( P_R )
  = D^{ \reg}_{ \sigma \mu }  ( T_k P_R )
 = { \chi^R ( T_k ) \over d_R }  D^{ \reg}_{ \sigma \mu }  (  P_R )
 \eea
 Using again Lemma 4.2 of \cite{TangQC}, then the error 
$\Delta ( \langle  X , Y \rangle ) = \epsilon ||X|| ||Y||$
can  be achieved with query complexity $ O ( { 1 \over \epsilon^2} ) $ 
(neglecting the failure probability $ \delta$).
As $ D^{ \reg}_{ \sigma \mu }  (  P_R )$ is queried by an oracle,  
 we have  
\bea 
\Delta (  \langle X , Y \rangle )  = \Delta (     { \chi^R ( T_k ) \over d_R }    )
 D^{ \reg}_{ \sigma \mu }  (  P_R )
\eea
Dividing by $ D^{ \reg}_{ \sigma \mu }  (  P_R )$,  one gets
\bea
\Delta (  {  \langle X , Y \rangle  \over   D^{ \reg}_{ \sigma \mu }  (  P_R ) }   ) 
= \Delta (     { \chi^R ( T_k ) \over d_R }    ) 
\eea
We want this uncertainty to be less than $1$: then, by the same
routine that ensures us  that we can use the classical algorithm to detect the normalized character, we set
\bea 
\Delta (     { \chi^R ( T_k) \over d_R }    )  = \epsilon ||X|| ||Y|| \le  1 
\eea
and obtain, for $k \ll n$ 
\bea 
\epsilon \le { 1 \over ||X|| ||Y|| }  =   \sqrt { n! \over d_R^2 } .  \sqrt{ k \over n (  n -1)\dots (n-k+1) }  \sim 
\sqrt {n! \over d_R^2 } \sqrt{ k \over n^{k} } 
\eea
Making use of the upper bound $\epsilon^* $ on $ \epsilon$, 
we write a lower bound on the query complexity which we call $Q_k^*$
\bea 
Q_k^* ( R )  = { 1 \over ( \epsilon^* )^2} = \frac{n^k}{k} . { d_R^2 \over (n!) } 
\eea
which holds for $k=2,3,\dots, k_*(n)$. 
Taking the worst case, this complexity is maximised for the largest $d_R$.
Keeping in mind  $(d_R^{\max})^2  \le  ( n! -2 )  $ \eqref{dRmax}, we are led to 
 \bea 
Q^*_k( R_{ max} ) \le \frac{n^k}{k} . { ( n! -2 )   \over (n!) }  
\sim \frac{n^k}{k} 
\eea
This is the behavior of the largest query complexity for each $k$. 
The total query complexity is obtained by summing them: 
\bea
 && 
 Q^*( R_{ max} ) = \sum_{k=2}^{k_*(n)}
Q^*_k( R_{ max} ) =  \sum_{k=2}^{k_*(n)}  \frac{n^k}{k} 
\geq {  n^{k_*(n)} \over  k_*(n) } 
= n^{n^\alpha - \alpha}
\eea
where  we assume $ k_*(n) = \cO (  n^{\alpha}) $ at large $n$ (for some $ 0 < \alpha < 1/2$, with $ \alpha = 1/4$ being a conjecture as discussed in section \eqref{sub:centre} ).   This  
shows that the query complexity of this algorithm is much larger than 
that of the quantum algorithm with $n^{2\alpha} \log n$, see \eqref{querysum}.  
%

\ 

\

%

\subsection{ Discussion and future directions }

\subsubsection{ Interpretation } 

We are finding, for the Young diagram projector detection task, that the quantum phase estimation methods perform exponentially better than the classical  analog we have defined using randomised algorithms (compare for example \eqref{tT2} with \eqref{ClassCompT2}).  This may be due to the intrinsically quantum nature of our task which may bring it closer to the exponential quantum improvements envisioned in \cite{HHL}. There is no computational cost of converting classical vector data to quantum states in our discussion of section \ref{sect:qpesn}. In practical recommendation tasks it has been found that the randomised algorithms only differ polynomially in complexity from the quantum tasks  \cite{Aaronson2015}\cite{TangQC}.


\subsubsection{One dimensional representations } 

We have asserted, in arriving at \eqref{bds},  that the one-dimensional representations  of the symmetric groups are the trivial and the sign representations. That there are no further one-dimensional irreps over $\mC$  for $ S_n $ ( $n \ge 2$)  is discussed in \cite{2784937} \cite{onedim1} \cite{onedim2}.

\subsubsection{ More general choices of $V$.   } 
We have formulated the randomised classical algorithm in terms of the regular representation. We could also use a more general reducible representation which contains all irreps $R$ in its decomposition into irreducibles. 
If $V$ contains every irrep $R$, then its dimension is at least 
\bea
\sum_{ R } d_R 
\eea
It is known that this sum is equal to the number of permutations $ \sigma $ which square to $1$. This means that $\sigma $ only has one and two-cycles. Let $k$ be the number of 2-cycles. 
\bea 
\sum_{ R } d_R  = \sum_{ k = 0 }^{ \lfloor { n \over 2 } \rfloor } 
{  n! \over 2^k k! ( n - 2k ) !  } 
\eea
The summand is maximum when $ k = (n/2)  $. The value is then 
\bea 
{ n ! \over ( n /2 )! 2^{ n/2} } 
\eea 
A lower bound is therefore 
\bea 
D_- = { n ! \over ( n /2 )! 2^{ n/2} } 
\eea
An upper bound is 
\bea 
D_+ = (  n/2 ) D_-  
\eea

\section{ Detection of the Kronecker projector in $ \cK(n)$  } 
\label{Kron}

Permutation centralizer algebras $\cK(n)$  related to $ \mC ( S_n)$, which arise in the enumeration of tensor model observables 
and the computation of their correlators \cite{BenGeloun:2013lim,BenGeloun:2017vwn}  have been used to give constructive integer matrix algorithms for computing  Kronecker coefficients 
 \cite{BenGeloun:2020yau,BenGeloun:2020lfe}.  These algorithms are based on regarding $ \cK ( n ) $ as  Hilbert spaces. The dimensions of these Hilbert spaces grow as $n! $ at large $n$ (see \cite{BenGeloun:2021cuj} for all orders asymptotic formulae for these dimensions).  An important motivation for this paper has been to take advantage of known exponential improvements due to quantum algorithms \cite{HHL} to formulae efficient algorithms based on $\cK ( n)$ for Kronecker coefficients, since  an important element of the interest in Kronecker coefficients comes from computational complexity theory 
\cite{Mulmuley2001GeometricCT, Burgisser2011NonvanishingOK,
Ikenmeyer2017OnVO, Pak2019OnTL}. Our perspective is that, while the intersection of representation theory in mathematics and computational complexity in computer science  is a fascinating interface with many interesting outcomes,  the mathematical question of combinatorial constructibility of representation theory quantities  is sometimes  fruitfully kept distinct from the complexity theoretic questions associated with the precise  choice of computational  task based on Kronecker coefficients and the efficiency of algorithms for the chosen task. Informally speaking we would argue that construction problems, such as  making hay and  manufacturing needles,  are distinct from  complexity questions,  such as the difficulty of  finding a needle in a  haystack.  We review the basics of $ \cK ( n )$ from \cite{BenGeloun:2017vwn}  and apply QPE to projector detection task in $\cK (n)$. The projector is labelled by a triple of Young diagrams with non-vanishing coefficient. 

Consider $ \mC ( S_n ) \otimes_{ \mC }  \mC ( S_n )$,  simply denoted
$ \mC ( S_n ) \otimes \mC ( S_n ) $. Introduce  the tensor product  elements
$ \sigma_1 \otimes \sigma_2 $ and sum over all their diagonal conjugates as 
\be \label{invgam}
\sigma_1 \otimes \sigma_2 \rightarrow \sum_{ \gamma \in S_n } \gamma \sigma_1 \gamma^{-1} \otimes 
\gamma \sigma_2 \gamma^{-1} 
\ee
Now, consider the $\mC$-vector subspace $\cK(n) \subset  \mC ( S_n ) \otimes \mC ( S_n )$ 
spanned by all $\sum_{ \gamma \in S_n } \gamma \sigma_1 \gamma^{-1} \otimes  \gamma \sigma_2 \gamma^{-1} $, 
$\s_1$ and $\s_2 \in S_n$: 
\be
\cK(n) = {\rm Span}_{\mC}\Big\{    
\sum_{ \gamma \in S_n } \gamma \sigma_1 \gamma^{-1} \otimes  \gamma \sigma_2 \gamma^{-1} , \; \s_1, \s_2 \in S_n
\Big\}
\label{graphbasis}
\ee
The dimension of $\cK(n) $ maps to the number of ribbon graphs with 
$n$ edges, and equivalently to a sum over triples $ R_1 , R_2 , R_3 $  of irreducible  representations (irreps)  of $S_n$  \cite{BenGeloun:2017vwn}: 
\bea 
|\Rib ( n ) | = \Dim ( \cK ( n ) ) = \sum_{ R_1 , R_2 , R_3 \vdash n } C ( R_1 , R_2 , R_3 )^2 
\eea
The Kronecker coefficient $ C ( R_1 , R_2 , R_3 )$ is a non-negative-integer 
that yields the number of times $R_3$ appears in the tensor product decomposition $ R_1 \otimes R_2$. The Kronecker coefficient is expressed in terms of characters  as 
\bea\label{CMN}  
C ( R_1  , R_2 , R_3 ) = { 1 \over n! } \sum_{ \sigma \in S_n } \chi_{ R_1 } ( \sigma ) \chi_{ R_2 } ( \sigma ) \chi_{ R_3 } ( \sigma ) 
\eea 

In an analogous way of section \ref{sect:qpesn}, there are operators
that multiplicatively generate the centre of $\cK(n)$. 
At any $n \ge 2$,  we define elements in   $ \mC ( S_n) \otimes \mC ( S_n)$  
\bea\label{tki}
T^{(1)}_k &=& T_k \otimes 1   = \sum_{ \s \in \cC_k } \sigma \otimes 1 \,,  \crcr
T^{(2)}_k &=& 1 \otimes T_k    = \sum_{ \s \in \cC_k } 1 \otimes \sigma \,, \crcr
T^{(3)}_k &=& \sum_{ \sigma \in \cC_k }  \sigma \otimes  \sigma  \;. 
\eea
The sum of products of the $T^{(i)}_k$'s, $k=1, \dots, n$, generates the centre
$\cZ (\cK(n))$ of $\cK(n)$. In fact, one does not need the entire set $k=1, \dots, n$
to generate the centre, only a fewer number of them is enough $k=1, \dots, k_*(n) \le n$. 

$\cK(n)$ has a matrix basis 
\bea 
&& Q^{ R_1 , R_2 , R_3 }_{ \tau_1 , \tau_2 } = {  d_{ R_1} d_{ R_2} \over (n!)^2 } 
\sum_{\tau = 1}^{C(R_1 , R_2 , R_3)}
\sum_{\s_1, \s_2 \in S_n}
\sum_{i_1,i_2,i_3, j_1,j_2} \cr 
&& ~~~~~~~~~~~~~~~~~~~~~~~~~~
C^{R_1,R_2; R_3 , \tau_1  }_{ i_1 , i_2 ; i_3 } C^{R_1,R_2; R_3, \tau_2  }_{ j_1 , j_2 ; i_3 } 
 D^{ R_1 }_{ i_1 j_1} ( \sigma_1  ) D^{R_2}_{ i_2 j_2 } ( \sigma_2 )  \,  \sigma_1 \otimes \sigma_2 
 \label{Qmunu}
\eea
obeying 
\bea 
Q^{ R_1 , R_2 , R_3 }_{ \tau_1 , \tau_2 } Q^{ R_1' , R_2' , R_3' }_{ \tau_1' , \tau_2' }
= \delta^{ R_1 , R_1'} \delta^{ R_2 , R_2'} \delta^{ R_3 , R_3'} 
\delta_{ \tau_2 , \tau_1' } Q^{ R_1 , R_2 , R_3 }_{ \tau_1 , \tau_2' }
\eea
 $D^R_{ ij} ( \sigma ) $ are the matrix elements of 
the linear operator $D^R(\s)$ in an orthonormal basis for the irrep $R$. The index
$ \tau $ runs over an orthonormal basis for the  multiplicity space 
of dimension equals to the Kronecker coefficient $C ( R_1 , R_2 , R_3 )$. 
$\kappa_{R_1,R_2, R_3}$ is a normalization factor. 
$C^{R_1,R_2; R_3 , \tau_1  }_{ i_1 , i_2 ;i_3 } $ are Clebsch-Gordan coefficients of the representations of $S_n$ (see the appendices 
of \cite{BenGeloun:2017vwn} for the properties needed to prove that this expression gives a Wedderburn-Artin basis for $ \cK(n)$). 

The centre  $\cZ (\cK(n))$ is spanned by 
\bea 
P^{ R_1 , R_2 , R_3 } = \sum_{ \tau }  Q^{ R_1 , R_2 , R_3 }_{ \tau , \tau } \nnm 
\eea
This is equal to an expression in terms of characters 
\bea 
\tilde P^{ R_1 , R_2 , R_3 }  && = \Delta ( P_{ R_3} )  ( P_{ R_1 } \otimes P_{ R_2 }  ) \cr 
&& = { 1 \over (n!)^3 } d_{ R_1 } d_{ R_2} d_{ R_3 } \sum_{ \sigma_1 , \sigma_2 , \sigma_3 \in S_n } 
\chi^{R_1 } ( \sigma_1 )    \chi^{R_2 } ( \sigma_2 ) \chi^{ R_3 } ( \sigma_3 ) ~~~
\sigma_3 \sigma_1 \otimes \sigma_3 \sigma_2
\label{Ptilde}
\eea
The equality $P^{ R_1 , R_2 , R_3 } = \tilde  P^{ R_1 , R_2 , R_3 } $  can be shown using the properties : 
\bea 
 P^{ R_1' } \tilde P^{ R_1 , R_2 , R_3 } &=&  \delta^{ R_1 , R_1'}  \tilde P^{ R_1 , R_2 , R_3 }  \cr 
 P^{ R_2' } \tilde P^{ R_1 , R_2 , R_3 } &=&  \delta^{ R_2 , R_2'} \tilde P^{ R_1 , R_2 , R_3 } \cr 
 P^{ R_3' } \tilde P^{ R_1 , R_2 , R_3 } &=&  \delta^{ R_3 , R_3'} \tilde P^{ R_1 , R_2 , R_3 }  \cr 
 \tilde P^{ R_1 , R_2 , R_3  }  \tilde P^{ R_1' , R_2' , R_3 ' }  
&=& \delta^{ R_1 , R_1'} \delta^{ R_2  , R_2' } \delta^{ R_3  , R_3' } 
   \tilde P^{ R_1 , R_2 , R_3  } 
\eea
It is known that, as $ ( R_1 , R_2 , R_3 ) $ range over triples of Young diagrams with non-vanishing Kronecker coefficient, and $\tau_1, \tau_2$ each range over a basis for the Kronecker multiplicity space, the elements  $ Q^{ R_1 , R_2 , R_3 }_{ \tau_1 , \tau_2 }  $ form a basis for $\cK ( n ) $ \cite{BenGeloun:2020yau}.  The above equations imply that 
\bea 
\tilde P^{ R_1 , R_2 , R_3 } = \sum_{ \tau } a_{ \tau } Q^{ R_1 , R_2 , R_3 }_{ \tau , \tau }
\eea
Further we show, in Appendix \ref{pfpropKron}, that for any fixed $ \tau $, 
\bea 
\tilde P^{ R_1 , R_2 , R_3 } Q^{ R_1 , R_2 , R_3 }_{ \tau , \tau } = Q^{ R_1 , R_2 , R_3 }_{ \tau , \tau } 
\label{PQ=Q}
\eea
Along with the use of the inner product on $\cK ( n) $ (described in \cite{BenGeloun:2020yau}) this equation implies  that $ \tilde P^{ R_1 , R_2 , R_3 }  = P^{ R_1 , R_2 , R_3 }$. 


\

\noindent 
{\bf SET-UP: }  Bob sends to Alice a non-vanishing projector $P^{ R_1 , R_2 , R_3 } $ associated with a triple $(R_1 , R_2 , R_3 )$ with non-vanishing Kronecker coefficient. Alice is  asked to detect the triple of Young diagrams labelling the projector. 

Alice uses  successive phase estimations with $U_{ k , i } = e^{ \frac{2 \pi i}{\chim_k}  T_k^{ (i) } } $, for $i \in \{ 1,2, 3 \}$ and $k \in \{ 1, 2, \cdots , k_* (n) \}$. For each $k, i $ we have a certain number of black-box queries in QPE ( for the controlled $U$-gates) and a number of gates for inverse quantum Fourier transformation.  We add up all the gates to get the gate complexity and all the black box queries to get the query complexity. 

\

\noindent {\bf TASK: }  {\bf Identify a triple $(R_1, R_2,R_3)$ for  a Kronecker  projector  $P^{R_1,R_2,R_3}$ }

Consider the set of unitaries $ U_{k,i}  = e^{  \frac{2 \pi i}{\chi^{\max}_k }   T_k^{(i)} } $, 
for any $i$ and $k$,  
that satisfy $ U_{k,i} P^{R_1,R_2,R_3} = e^{  \frac{2 \pi i}{\chi^{\max}_k }   \widehat {\chi}^{R_i} ( T_k^{(i)} ) }  P^{R_1,R_2,R_3}$.  The problem is formulated as follows: 

\begin{center}
``Given the set of unitaries $U_{2,i}, U_{3,i}, \dots, U_{k,i}$, $i=1,2,3$, $k\in [ \! [2, k_*(n)  ]\! ]$, and state $P^{R_1,R_2,R_3}$ with unknown $R_i\vdash n$, $i=1,2,3$, determine which $P^{R_1,R_2,R_3}$ we have.''
\end{center}

The reasoning is analogous of the previous section  \ref{sect:qpesn}. 
The number of bits require to store de top eigenvalue  of $T_k^{(i)}$ 
for any $i$  and $k$ is known \eqref{bittk}. The query complexity 
adds the number of queries of all C-$U_{k,i}$-operators
\bea
\sum_{i=1}^3 
\sum_{ k =2}^{ k_* (n) }   k \log (n ) = 3 \Big( {  k_* (n) ( k_*(n) +1 )\over 2 } - 1 \Big)  \log (n ) 
\eea
Using again $k_*(n) = n^{\alpha}$, we obtain 

\noindent{\bf Query complexity} $ \in \cO(t) \sim \cO(n^{2\alpha} \log n)$. 

In a similar way, the gate complexity  sums the number of gates required for each step and estimate
\bea 
\sum_{i=1}^3 
\sum_{ k =2}^{ k_* (n) }   k^2 ( \log (n ) )^2  \sim    k_* (n)^3  ( \log (n ) )^2    
\eea
If we use $k_*(n) = n^{ \alpha } $ and at large $n$, this boils down to 

\noindent{\bf Gate complexity} $ \in  \cO(n^{3\alpha} (\log n)^2 )$.

\vskip.2cm

\noindent
{\Large {{\bf Future directions } }} 

\vskip.2cm 

\noindent 
{\bf   Generating list of triples with non-vanishing Kronecker coefficients} 

\vskip.1cm

We have presented complexity estimates for the detection of Kronecker projectors which exist for every triple of Young diagrams having non-vanishing Kronecker coefficients. A natural next step in understanding Kronecker coefficients in the context of ribbon graph quantum mechanics is to consider the ribbon graph corresponding to state in $\cK ( n ) $ corresponding to the permutation pairs $ ( \sigma_1 = \sigma_0  , \sigma_2 = \sigma_0 )  $ 
where $ \sigma_0 $ is the identity permutation. When this is expanded in the Fourier basis, we get a linear combination with non-vanishing coefficients of  the projector states $P^{ R_1 , R_2 , R_3 } $ for 
all the triples $(R_1, R_2, R_3 )$ having non-vanishing Kronecker coefficients. The application of  QPE with the operators  $U_{ k , i } $ will be able to produce the list of Young diagram triples with non-vanishing Kroneckers. We leave the complexity estimates for this task to the future.

\vskip.2cm 

\noindent 
{\bf Detecting whether the Kronecker coefficient for a triple is greater than $1$ } 

\vskip.1cm 

By focusing on the detection of the projector states $P^{ R_1, R_2, R_3 }$ in $\cK(n ) $ we have explained algorithms for identifying/detecting specific  triples from the set of all triples with non-vanishing Kronecker coefficient. Making use of  the more refined  structure of $\cK ( n)$, notably its Wedderburn-Artin decomposition in terms of $Q^{ R_1 , R_2 , R_3 }_{ \tau_1 , \tau_2 } $   would allow access to generalisations of this task. 
For example Bob could send Alice a pair of  $Q$-states, both with the same $( R_1, R_2, R_3) $ and with Kronecker coefficient $1$ or greater than $1$,   and Alice could be tasked with determining the triple of Young diagrams and determining whether the two states sent by Bob are linearly independent. If they are, then this would show that the Kronecker coefficient for the triple is greater than $1$. This would require using QPE to determine the eigenvalues of elements in $\cK ( n ) $ which are not necessarily central. The use of non-central elements of a permutation centraliser algebra to construct distinct multiplicity states $(\tau_1, \tau_2 )$ has been used \cite{Kimura:2008ac}  in  the context of multi-matrix invariants (related to the algebra $\cA ( m ,n )$ discussed in  section \ref{LR}).

\vskip.2cm 

\noindent 
{ \bf  Holography  } 

\vskip.1cm 
Following the similarity between the projector detection in this section and the detection in $ \cZ ( \mC ( S_n) )$ from section \ref{sect:qpesn} it is natural to ask if we can formulate a holographically  dual gravitational detection problem for $\cK ( n)$. This algebra has been studied \cite{Mattioli:2016eyp}\cite{BenGeloun:2017vwn}  in connection with invariant observables for tensor models. These in turn have been related \cite{Witten:2016iux} to the SYK models \cite{SY1993,Kitaev}. A holographic correspondence between SYK and near-$AdS_2$ quantum gravity has been discussed (see the original papers \cite{Polchinski:2016xgd,Jevicki:2016bwu,Maldacena:2016upp} and a recent review which discusses the correspondence and associated subtleties \cite{Sarosi:2017ykf}).  It is an interesting question whether near-AdS2 gravity allows a classical gravity dual of the projector detection in $\cK ( n)$ analogous to the classical gravity dual of projector detection in $  \cZ ( \mC ( S_n ))  $ based on LLM geometries \cite{Lin:2004nb} which we described in  section \ref{holog}.

\section{Detection of the Littlewood-Richardson projector} 
\label{LR}

Given two non negative integers $m$ and $n$, 
there is a  PCA $\cA ( m , n ) $  which  is relevant for the counting of polynomial functions of two matrices $X , Y $ invariant under adjoint action by unitary matrices $U :  ( X , Y ) \rightarrow ( U X U^{ \dagger} , U Y U^{ \dagger}  ) $.   We present a review of the background from  \cite{Mattioli:2016eyp}. 

$\cA(m,n)$ is the subalgebra of $\mC(S_{ m + n })$ made of the elements that
are invariant under the adjoint action of $S_{ m}  \times S_n$. 
There is a basis of  $ \cA(m,n)$ labelled by multi-necklaces with $m$ beads of one colour and $n$ beads of another colour. 
These are in 1-1 correspondence with orbits in $S_{m+n}$ generated by the conjugation action of $\gamma \in S_{ m } \times S_n \subset S_{ m+n}$.  With $r$ running over these necklaces, 
\bea 
E_r = \sum_{\g \in S_{ m}  \times S_n}  \g \s^{(r)}  \g^{-1}, 
\eea
where $\s^{(r)}  \in S_{ m + n } $ is a representative of the $r$'th orbit.   
One shows that $\cA(m,n)$  is a semi-simple associative algebra 
and therefore admits a Wedderburn-Artin decomposition. 
It is proved that
\bea
\Dim (\cA(m,n)) = \sum_{R \vdash m+n, R_1 \vdash m, R_2 \vdash n} 
g(R_1, R_2, R)^2 
\eea
where $g(R,R_1, R_2)$ is the so-called Littlewood-Richardson (LR) coefficient. 

There is a Wedderburn-Artin basis of $\cA( m ,n )$ in the form  $Q^{ R}_{ R_1 , R_2 ; \mu  \nu }$
with $R , R_1 , R_2 $ being Young diagrams with $m+n, m , n $ boxes respectively. The Littlewood-Richardson coefficient determines the range of the indices $\mu, \nu$ :  $1 \le \mu , \nu \le g( R_1 , R_2, R )  $. 
Explicit formulae for $Q^{ R}_{ R_1 , R_2 ; \mu  \nu }$ are known  in terms of 
matrix elements  of permutations in the irrep $R$ of $S_{m+n}$ along with branching coefficients for the reduction of the irrep $R$ into representations of the subgroup $S_m \times S_n$. The centre $\cZ (\cA(m,n))$ is spanned by projectors  labelled by triples $( R_1 , R_2 , R_3 ) $ with non-vanishing $g ( R_1 , R_2 , R ) $. They can be written in terms of characters : 
\bea 
P^{ R }_{ R_1 , R_2 }  = { d_R d_{ R_1} d_{ R_2} \over ( m+n)! m! n! } \sum_{ \sigma \in S_{ m+n} } 
\sum_{  \sigma_1 \in S_m } \sum_{ \sigma_2 \in S_n } \chi_R ( \sigma ) \chi_{ R_1} ( \sigma_1) \chi_{ R_2} ( \sigma_2 )  ~~ \sigma ( \sigma_1 \circ \sigma_2) 
\eea

\ 

\noindent {\bf TASK: }  {\bf Identify a triple $(R, R_1, R_2)$ for  a LR-projector  $P^{R}_{R_1,R_2}$ }

We proceed again by making explicit our set of unitaries that
will serve in the QPE of this system.  For all $n$, we introduce the notation 
$T^{(S_n)}_\mu$  which keeps its meaning of \eqref{Tmu}, but  use the superscript 
$S_n$ as we are dealing with different symmetric group in the present formalism. 
As usual $\mu$ will be restricted to $[k,1^{ n-k}]$. 
We also introduce 
\bea
 \chi^{\max}_{n, k } = \max_{R \vdash n} \widehat \chi^R ( T^{(S_n)}_{ k })
\eea
and  define 
 the unitary operators
\bea 
&& U_{m,  k }^{(1)} = e^{ \frac{2 \pi i}{\chimm} T^{ (S_m)}_{ k} } \otimes 1 \cr 
&& U_{n,k}^{ (2) } = 1 \otimes e^{  \frac{2 \pi i}{\chimn} T_k^{ (S_n)}  } \cr 
&& U_{ m+n, k }^{ (3)} = e^{  \frac{2 \pi i}{\chimp} T_k^{ (S_{ m+n} ) }}
\eea
There operators satisfy the eigenvalue equations 
\bea
&&
 U_{m, k}^{(1)}  P^{R}_{R_1,R_2}= e^{ \frac{2 \pi i}{\chi^{\max}_{n,k} } \widehat \chi^{R_1} ( T^{(S_m)}_{ k} )  }   P^{R}_{R_1,R_2}, \crcr
  && 
 U_{n, k}^{(2)} P^{R}_{R_1,R_2}  = e^{  \frac{2 \pi i}{\chi^{\max}_{n,k} }  
 \widehat \chi^{R_2}( T^{(S_n)}_{ k }) } P^{R}_{R_1,R_2} , \crcr
&&
 U_{m+n, k}^{(3)}  P^{R}_{R_1,R_2} = e^{  \frac{2 \pi i}{\chi^{\max}_{m+n,k} }  \widehat \chi^R ( T^{(S_{n+m})}_{ k}) } 
  P^{R}_{R_1,R_2}
\eea
for any $k$. This will play an important role in the QPE formalism. 
We need to use the values of 
\bea 
&&  k_1 \in \{ 2, 3, \cdots , k_* (m ) \} \cr 
 && k_2 \in \{ 2, 3, \cdots , k_* (n) \} \cr 
 && k_3 \in \{ 2, 3 , \cdots , k_* ( m+n ) \} 
\eea 
  to detect the Young diagrams $R_1 , R_2 $ and $R$,  respectively. 

The problem of detecting LR-projectors is formulated as follows: 

\begin{center}
``Given the set of unitaries 
$U^{(i)}_{\ell_i,k_i  }, U^{(i)}_{\ell_i ,k_i }, \dots,  U^{(i)}_{\ell_i, k_i} $, $i=1,2,3$, 
$\ell_1= m $, $\ell_2= n$, $\ell_3=m+n$, 
$k_i\in [ \! [2, k_*(\ell_i)  ]\! ]$, and state $P^{R}_{R_1,R_2}$ with unknown $R_i\vdash n_i$, $i=1,2$,
$R\vdash m+n$, determine which  $P^{R}_{R_1,R_2}$ we have.''
\end{center}

To obtain the query and gate complexities it suffices to use the same reasoning
as previously. The difference  in this setting is that they 
becomes function of the pair $(m,n)$. 
The number of bits required to store the different eigenvalues
of the operators are listed below: 
\bea \label{bittk}
&&
\hbox{ Bits required to store eigenvalues of $T^{(S_m)}_{ k_1}$  } 
 \sim  t_1 \sim   k_1 \log m    \crcr
 &&
 \hbox{ Bits required to store eigenvalues of $T^{(S_n)}_{ k_2}$  } 
 \sim  t_2 \sim   k_2 \log n   \crcr
 &&
  \hbox{ Bits required to store eigenvalues of $T^{(S_{n+m})}_{ k_3}$  } 
 \sim  t_3 \sim  k_3 \log (n +m)  \cr 
 && 
\eea
We add up  the complexities for the measurements of in each sector
and find its estimate:  
\bea
&&
\sum_{ k_1 =2}^{  k_*(m)  }   k_1  \log (m ) \sim   k_* (m)^2  \log (m )  \,, 
\qquad 
\sum_{ k_2 =2}^{ k_* (n) }   k_2 \log (n ) \sim    k_* (n)^2  \log (n )   \crcr
&&
\sum_{ k_3 =2}^{ k_* (n+m) }   k_3 \log (m+n )\sim    k_* (m+n)^2  \log (m+n)
\eea
If   $ k_* (p) < p^{ \alpha } $ holds at large $p$, we have 
\bea
&&
t \sim   (m+n)^{2\alpha}  \log (m+n)
\crcr
&&
\eea
and therefore

\noindent{\bf Query complexity} $ \in \cO(t) \sim \cO( (m+n)^{2\alpha}  \log (m+n))$. 

The gate complexity is estimated in the same way: 

\noindent{\bf Gate complexity} $ \in \cO( (m+n)^{3\alpha}  (\log (m+n))^2 )$.

\vskip.3cm

\centerline{\bf{Acknowledgements}}
\vskip.2cm 

We are pleased to thank Stephon Alexander, George Barnes,  Robert de Mello Koch, Humberto Gilmer, Antal Jevicki, Caroline Klivans, Yangrui Hu,  Garreth Kemp, David Lowe, Tucker Manton and Adrian Padellaro for insightful discussions during the course of the project. SR is supported by the STFC consolidated grant ST/P000754/1 ``String Theory, Gauge Theory and Duality''.  The authors acknowledge support of the Institut Henri Poincaré (UAR 839 CNRS-Sorbonne Université), and LabEx CARMIN (ANR-10-LABX-59-01).  SR acknowledges the theoretical physics group at Brown University and the Perimeter Institute for hospitality during the final stages of completion of this work.

\vskip.5cm 

\begin{appendix}

\section{ Representation theory lemmas used in quantum-inspired classical algorithms } 
\label{app:prooflem}

This appendix provides the proof of two lemmas used in 
section \ref{sect:randclass} and further remarks. 

\

\noindent 
{\bf  Proof of Lemma \ref{lem1} } 
We start with the definition of the representation matrice element 
of a projector  
\bea 
 D^{ \reg}_{ \gamma \mu } ( P_R ) = \sum_{ \sigma } { d_R \chi_R ( \sigma ) \over n! } 
D^{ \reg}_{ \gamma \mu } ( \sigma ) 
 = \sum_{ \sigma } { d_R \chi_R ( \sigma ) \over n! } \delta ( \gamma^{-1} \sigma \mu ) 
\eea
Taking the sum of squares, we obtain 
\bea 
&& \sum_{ \gamma} ( D^{ \reg}_{ \gamma \mu } ( P_R ) )^2 \cr 
&& = \sum_{ \gamma  } { 1 \over (n!)^2 } \sum_{ \sigma_1} d_R \chi^R ( \sigma_1 ) \delta ( \gamma^{-1} \sigma_1 \mu ) \sum_{ \sigma_2} d_R \chi^R ( \sigma_2 ) \delta ( \gamma^{-1} \sigma_2 \mu ) \cr 
&& = \sum_{ \sigma_1 , \sigma_2 } d_R^2  \chi^R ( \sigma_1 ) \chi^R ( \sigma_2 ) 
\delta ( \mu^{-1} \sigma_1^{-1} \sigma_2 ) \mu ) \cr 
&& = \sum_{ \sigma_1 . \sigma_2 } { d_R^2 \over ( n! )^2 } \chi^R ( \sigma_1 ) \chi^R ( \sigma_2 ) \delta ( \sigma_1^{-1} \sigma_2 ) \cr 
&& = \sum_{ \sigma_1 } { d_R^2 \over ( n!)^2 } \chi^R ( \sigma_1 ) \chi^R ( \sigma_1 ) \cr 
&& = { d_R^2 \over n! } 
\eea
{\bf Remark.} This is independent of the choice of $\mu $ and is the Plancherel probability weight for the irrep $R$. 

\noindent
{ \bf Proof of Lemma \ref{lem2} } \\
\bea 
 \sum_{ \tau }   ( D^{ \reg}_{ \sigma \tau }  ( T_2 ) )^2 & =& \sum_{ \tau } \delta ( \sigma^{-1} T_2 \tau ) \delta ( \sigma^{-1} T_2 \tau )  \cr 
& =& \delta ( \sigma^{-1} T_2^2 \sigma ) \cr 
& =& \delta ( T_2^2 ) = { n ( n -1 ) \over 2 } 
\eea

\section{Solving the power sums $M_k$}
\label{solvingM}
The problem of solving the power sum moments $M_k= \sum_{i=1}^N f_i^k$ can be sorted by a
simple resolution of  a  linear system that we now present. 

We equate 
\bea
 \sum_{ f\in \cF } A^{ l } ( f ) =  \sum_{k=0}^l  c^l_k M_k 
\eea
and   expand the l.h.s using equation (3.9) in \cite{Balasubramanian:2006jt}
\bea
&&
 \sum_{ f\in \cF } A^{ l } ( f )  = 
 \sum_{i=1 }^N  A^{ l } ( f_i ) = 
  \sum_{i=1 }^N  l! \sum_{r=0}^l \binom{l}{r}\binom{f_i}{r} 2^r 
  \crcr
  &&
  = 
 \sum_{r=0}^l   \frac{ 2^r  l!}{r!} \binom{l}{r}  \sum_{i=1 }^N (f_i)_r
  = 
   \sum_{r=0}^l   \frac{ 2^r  l!}{r!} \binom{l}{r}  \sum_{i=1 }^N 
   \sum_{k=0}^r s(r,k) f_i^k 
\eea
where $(f_i)_r = f_i(f_i - 1) \dots (f_i-r+1)$ is the failing factorial, 
and $s(k,r)$'s are Stirling numbers of the first kind. 
This expands further  as
\bea
 \sum_{ f\in \cF } A^{ l } ( f )  = 
   \sum_{k=0}^l 
\left[   \sum_{r=k}^l   s(r,k) \frac{ 2^r  l!}{r!} \binom{l}{r}
\right]
   \Big(  \sum_{i=1 }^N   f_i^k \Big) 
    = \sum_{k=0}^l 
\left[   \sum_{r=k}^l   s(r,k) \frac{ 2^r  l!}{r!} \binom{l}{r}
\right] M_k 
\label{sumA}
\eea
Thus the coefficient $c_k$ is 
\bea
c_k^l = \sum_{r=k}^l   s(r,k) \frac{ 2^r  l!}{r!} \binom{l}{r}
\eea
Fixing $l=0,1,\dots, \Lambda$, and having sorted 
$U(l,\rho) =  (-1)^l (l+1)\sum_{ f\in \cF } A^{ l } ( f ) /\rho^{2l+2}$,  
we can invert
the above triangular system \eqref{sumA} starting from $l=0$. 

 
\section{ Proof of a property of Kronecker projectors } 
\label{pfpropKron}

This appendix proves the relation \eqref{PQ=Q}.  Using 
$\tilde P^{ R_1, R_2 , R_3 }$ in \eqref{Ptilde}, 
$Q^{ R_1' , R_2' , R_3' }_{ \mu \mu } $ in \eqref{Qmunu}, 
and introducing the notation 
$\kappa_{R'_1R'_2} =  {  d_{ R'_1} d_{ R'_2} \over (n!)^2 } $, 
we have 
\bea 
&& \tilde P^{ R_1, R_2 , R_3 } Q^{ R_1' , R_2' , R_3' }_{ \mu \mu } 
=\kappa_{R'_1R'_2}   { d_{ R_1} d_{ R_2} d_{ R_3} \over (n!)^3 } 
\sum_{ \tau_1  , \tau_2 , \tau_3 }  \sum_{ \sigma_1 , \sigma_2 } \chi^{ R_1 }  ( \tau_1 ) \chi^{ R_2 } ( \tau_2 ) \chi^{ R_3 } ( \tau_3 ) 
\cr 
&& \qquad \qquad \qquad  \sum_{ i_1 , i_2 , j_1 , j_2 , k } 
D^{ R_1'}_{ i_1 j_1 } ( \sigma_1 ) D^{ R_2'}_{ i_2 j_2} ( \sigma_2 ) 
C^{ R_1' R_2' ; R_3' , \mu }_{ i_1 , i_2 , k }  C^{ R_1' R_2' ; R_3' , \mu }_{ j_1 , j_2 , k }  
~~~ \tau_3 \tau_1 \sigma_1 \otimes \tau_3 \tau_2 \sigma_2 \cr 
&& = \kappa_{R'_1R'_2}    \sum_{ \tau_1 , \tau_2  , \tau_3 } \sum_{ \tilde \sigma_1 , \tilde \sigma_2 }  { d_{ R_1} d_{ R_2} d_{ R_3 }  \over (n!)^3  } 
\chi^{ R_1 } ( \tau_1 ) \chi^{ R_2} ( \tau_2 ) \chi^{ R_3 } ( \tau_3 ) \cr 
&&  \qquad \qquad \qquad    \sum_{ i_1 , i_2 , j_1 , j_2 , k }  D^{ R_1'}_{ i_1 j_1 } ( \tau_1^{ -1} \tilde \sigma_{ 1} ) D^{ R_2'}_{ i_2 j_2 } ( \tau_2^{-1} \tilde \sigma_2 )C^{ R_1' R_2' ; R_3' , \mu }_{ i_1 , i_2 , k }  C^{ R_1' R_2' ; R_3' , \mu }_{ j_1 , j_2 , k }   \tau_3 \tilde \sigma_1 \otimes \tau_3 \tilde \sigma_2 \cr 
&& = \delta^{ R_1 R_1'} \delta^{ R_2 R_2' }\kappa_{R_1R_2}   \sum_{ \tau_3 }\sum_{ \tilde \sigma_1 , \tilde \sigma_2 }  { d_{ R_3 } \chi^{ R_3 } ( \tau_3 ) \over n!  } \cr 
&&\qquad \qquad \qquad 
\sum_{ i_1 , i_2 , j_1 , j_2 , k }  D^{ R_1 }_{ i_1 j_1 } ( \tilde \sigma_1 ) D^{ R_2 }_{ i_2 j_2 } ( \tilde \sigma_2 ) 
C^{ R_1 , R_2 ; R_3 ,\mu }_{ i_1 , i_2 , k } C^{ R_1 , R_2 ; R_3 , \mu }_{ j_1 , j_2 , k } 
( \tau_3 \tilde \sigma_1 \otimes \tau_3 \tilde \sigma_2 )  
\cr 
&& = \delta^{ R_1 R_1'} \delta^{ R_2 R_2'}\kappa_{R_1R_2}  \sum_{  \tau_3 } \sum_{ \sigma_1 , \sigma_2 } { d_{ R_3 } \chi^{ R_3 } ( \tau_3 ) \over n!  } 
 \cr 
&&\qquad \qquad \qquad 
\sum_{ i_1 , i_2 , j_1 , j_2 , k } 
D^{ R_1 }_{ i_1 j_1  } ( \tau_3^{ -1}  \sigma_1 ) D^{ R_2 }_{ i_2  j_2 } ( \tau_3^{ -1}  \sigma_2 ) 
C^{ R_1 R_2 R_3 ; \mu }_{ i_1 i_2 k } C^{ R_1 R_2 R_3 ; \mu }_{ j_1 j_2  k } \sigma_1 \otimes \sigma_2 \cr 
&& = \delta^{ R_1 R_1' } \delta^{ R_2 R_2' } \kappa_{R_1R_2} \sum_{ \sigma_1 , \sigma_2 } 
\sum_{ i_1 , i_2 , j_1, j_2 } D^{ R_1 }_{ i_1 p_1 } ( \tau_3^{-1} ) D^{ R_1 }_{ p_1 j_1 } ( \sigma_1 ) 
D^{ R_2}_{ i_2 p_2 } ( \tau_3^{ -1} ) D^{ R_2 }_{ p_2 j_2 } ( \sigma_2 ) 
\cr 
&&\qquad \qquad \qquad 
\times 
C^{R_1 , R_2  , R_3 ; \mu }_{ i_1 i_2 k }   C^{R_1 , R_2  , R_3 ; \mu }_{ j_1 j_2 k } \sigma_1 \otimes \sigma_2 \cr 
&& = \delta^{ R_1 R_1' } \delta^{ R_2 R_2'} \delta^{ R_3 R_3'}  \kappa_{R_1R_2} \sum_{ \sigma_1 , \sigma_2 } 
D^{ R_1 }_{ p_1 j_1 } ( \sigma_1 ) D^{ R_2 }_{ p_2 j_2 } ( \sigma_2 ) 
\cr 
&&\qquad \qquad \qquad 
\times 
C^{ R_1  , R_2 , R_3 }_{ p_1 , p_2 , k' } D^{ R_3' }_{ k' k } ( \tau_3 ) \delta^{ \nu\mu }  
{  d_{ R_3 } \chi^{ R_3 } ( \tau_3 ) \over n ! } C^{ R_1 R_2 ; R_3 , \mu }_{ j_1 , j_2 , k } \sigma_1 \otimes \sigma_2  \cr 
&& = \delta^{ R_1 R_1' } \delta^{ R_2 R_2'} \delta^{ R_3 R_3'} \kappa_{R_1R_2}    \sum_{ \sigma_1 , \sigma_2 } D^{ R_1  }_{ p_1 j_1 } ( \sigma_1 ) 
D^{ R_2 }_{ p_2 j_2 } ( \sigma_2 ) C^{ R_1 , R_2 ; R_3 , \mu }_{ p_1 , p_2 , k } 
C^{ R_1 , R_2 ; R_3 , \mu }_{ j_1 , j_2 , k }  \sigma_1 \otimes \sigma_2 \cr 
&& = \delta^{ R_1 , R_1'} \delta^{ R_2 , R_2' } \delta^{ R_3 , R_3' } 
Q^{ R_1 , R_2 , R_3}_{\mu \mu }  
\eea

\end{appendix}

\bibliographystyle{amsplain-ac}
\bibliography{mybib}

\end{document}